# Barium Titanate and Lithium Niobate Permittivity and Pockels Coefficients from MHz to Sub-THz Frequencies

Daniel Chelladurai[1*], Manuel Kohli[1], Joel Winiger[1], David Moor[1], Andreas Messner[1†], Yuriy Fedoryshyn[1], Mohammed Eleraky[2], Yuqi Liu[2], Hua Wang[2] and Juerg Leuthold[1]

[1]Institute of Electromagnetic Fields (IEF), ETH Zurich, Gloriastrasse 35, 8092 Zurich, Switzerland
[2]Integrated Devices, Electronics, And Systems (IDEAS) Group, ETH Zurich, Gloriastrasse 35, 8092 Zurich, Switzerland
[†]Now with: Zurich Instruments AG, 8005 Zurich, Switzerland
*Corresponding authors: daniel.chelladurai@ief.ee.ethz.ch, juerg.leuthold@ief.ee.ethz.ch

## Abstract

The Pockels effect is essential for controlling optical signals at the highest speeds, particularly for electro-optic modulators in photonic integrated circuits. Lithium niobate (LN) and barium titanate (BTO) are two excellent Pockels materials to this end. Here, we measure the Pockels coefficients and permittivity in LN and BTO over a continuous frequency range from 100 MHz to 330 GHz. These properties are constant across this frequency range in LN but have a significant frequency dependence in BTO. Still, our measurements show that BTO has remarkably large electro-optic properties compared to LN. Furthermore, we show how BTO devices can be designed with a flat electro-optic frequency response despite the Pockels coefficient dispersion. Finally, we expound our method for broadband characterization of these vital electro-optic properties, utilizing specialized integrated electro-optic phase shifters. Altogether, this work empowers the design of high-speed BTO devices and the development of new electro-optic materials.

## Introduction

The Pockels effect, or linear electro-optic (EO) effect, is a widely used nonlinear optics phenomenon where a near-instantaneous refractive index change is induced by applying an electric field. The fast response is ideal for optical communications where it enables EO modulator bandwidths to reach beyond 100 GHz[1–4]. Pockels materials are used in the active region of these modulators to map an electrical signal onto an optical carrier. Pockels modulators find use in applications at lower speeds, too. For example, they are useful as microwave-optical transducers in quantum networks[5,6], or as switches in programmable photonic circuits[7–9]. Pockels materials have also been demonstrated in reconfigurable metasurfaces[10,11] which may be interesting for LIDAR, augmented reality and virtual reality applications. This wide range of applications requires the Pockels effect to work across an accordingly wide range of operating frequencies spanning from MHz to 100s of GHz. It is therefore of high interest to know which EO materials can efficiently cover this range.

Ferroelectric materials like lithium niobate (LN) and barium titanate (BTO) are popular Pockels materials because their large optical refractive index is useful for waveguiding and their large Pockels coefficients are useful for EO effects. However, the frequency-dependence of the Pockels coefficients in these materials has not been experimentally verified to the best of our knowledge. Permittivity measurements are more commonly reported, yet measurements for thin films at frequencies above 10 GHz are scarce, especially for BTO. A simple and effective method for extracting the frequency dependent permittivity and EO coefficients of a thin film directly from a device in photonic integrated circuits is highly desirable.

In this work we experimentally verify the frequency responses of the permittivities as well as the Pockels coefficients for LN and BTO covering the range from 100 MHz to 330 GHz. To this end, we introduce a method to characterize the Pockels coefficient and permittivity of thin films using specialized phase shifters in photonic integrated circuits. Furthermore, by comparing the frequency dependence of BTO's permittivity to its Pockels coefficients, we show that the dispersion is linked to the $r_{42}$ Pockels coefficient and the permittivity along the crystalline $a$-axis $\varepsilon_a$. Meanwhile the $r_{33}$ coefficient and $\varepsilon_c$ remains constant above 1 GHz. We also present the interesting conclusion that for certain device geometries the EO response in BTO devices can be kept constant with frequency despite the strong dispersion in its EO properties.

### Device For Electro-Optic Characterization

The EO characterization of thin films up to sub-THz frequencies presents numerous challenges. Unlike bulk crystals, it is not easy to place thin films in the path of a free-space interferometer. Integrated photonic devices must be used instead. While integrated EO devices are not new, they are rarely operated at frequencies above 110 GHz because of bandwidth limitations. These bandwidth limitations arise from the RC time constant associated with charging the electrodes, RF attenuation along the length of the modulator and/or velocity mismatch between the RF and optical signals[12]. The latter two effects can be mitigated if the length of the device is sufficiently short. This, however, comes with the trade-off of reduced EO efficiency and an RC limit will still exist. Additionally, it is difficult to generate signals at sub-THz frequencies with powers that are high enough to distinguish weak EO effects from noise. To measure the weak Pockels shift in bandwidth-limited or short, inefficient devices, it is necessary to have high optical input powers combined with low optical insertion loss. With the above points in consideration, the ideal EO characterization device should have a short length, low optical insertion loss and high EO efficiency per unit length.

In pursuit of an ideal EO characterization device, we have developed a specialized phase shifter that is based on hybrid gap plasmon waveguides (HGPW)[13,14]. Compared to those previous demonstrations, we replace the silicon layer of the HGPW structure with an EO material. In this way, the optical mode is confined within an EO material even when the gap between the electrodes is large and the optical losses are small. A cross-sectional schematic of the device is shown in Fig. 1(a). A thin film of an EO material like LN or BTO on a buried oxide is the primary waveguiding layer. Fig. 1(b) shows the optical mode's electric field intensity profile at a wavelength of 1550 nm. The electrodes guide the optical mode similarly to nanoscale plasmonic slot modulators, but with a much larger spacing between the electrodes to reduce optical propagation losses. An electrical signal applied to the electrodes results in the RF electric field profile in Fig. 1(c). The presence of this RF electric field in the EO material causes a change in refractive index due to the Pockels effect according to equation (1)[15]

$$\Delta n = -\frac{1}{2} n_0^3 r_{\text{eff}} E \,, \tag{1}$$

where $n_0$ is the material's unperturbed refractive index, $r_{\text{eff}}$ is the effective Pockels coefficient and $E$ is the RF electric field in the material. Fig. 1(d) shows this spatially varying refractive index change $\Delta n$. The visualizations in Fig. 1 are qualitative. The values in each figure are normalized such that the maximum is unity. In reality, the true values depend on the optical and RF powers as well as the permittivity and Pockels coefficients, which can vary widely between EO materials.

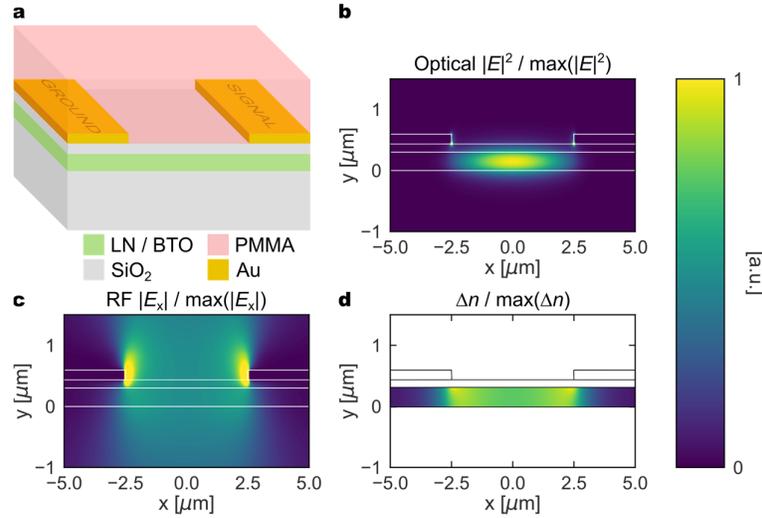

Fig. 1 | Integrated phase shifter used to measure permittivity and Pockels coefficients. a, Schematic cross-section of the phase shifter used for EO characterization. b, E-field profile of an optical mode in the phase shifter for $\lambda$ = 1550 nm. c, RF E-field profile when a 50 GHz electrical signal is applied to the electrodes. d, Spatially varying refractive index change in the EO material due to the Pockels effect. The maximum $\Delta n/\Delta V$ is on the order of $10^{-5}$ V$^{-1}$ for LN and $10^{-4}$ V$^{-1}$ for BTO. Note that the values in each of **b-d** are normalized to fit the same

There are several benefits to using this kind of device for EO characterization. Firstly, it covers the aspects of the ideal EO characterization device discussed above. The phase shifters are efficient. The optical field is mostly confined to the EO layer and, more importantly, also within the region that undergoes a change in refractive index. Since the devices are efficient, their lengths can be reduced to only 10s of microns (25 µm for BTO or 50 µm for LN) while still providing useful modulation. With such short devices, the capacitance is small and the RC bandwidth limitation is pushed to frequencies in the 100s of GHz. There is no need for impedance matching or RF terminations and the effects of a velocity mismatch, or walk-off, between the RF and optical signals is negligible. The short devices also have negligible on-chip insertion losses, which helps with measuring Pockels coefficients despite weak EO signals. The total insertion loss is approximately 0.2 dB based on mode simulations and supported by our previous experiments[16]. For a phase shifter with a 50 µm length and a 5 µm electrode separation, propagation losses account for roughly 0.1 dB (20 dB/cm) with the remaining 0.1 dB caused by reflection losses at the transitions to and from the device. The phase shifter also benefits from a straightforward fabrication process. As described in the methods section, both the structure and the fabrication process leave little uncertainty with respect to dimensions and alignment.

## Permittivity

There are stark differences between the permittivities of LN and BTO as a function of frequency. In this section we discuss the differences between thin films of the two materials and present measurements of their permittivities alongside available literature data. In Supplementary Note 1 we elaborate on the origins of the permittivity in ferroelectrics. The crystal structures of LN and BTO both lead to large anisotropies in their respective permittivity. Fig. 2(a) shows a sketch of the trigonal unit cell for LN belonging to the 3$m$ point group while Fig. 2(b) shows the same for BTO's tetragonal unit cell belonging to the 4$mm$ point group. For clarity, most of the oxygen atoms in the LN unit cell have been omitted. The direction of the displacement of cations along the $c$-axis determines the direction of the spontaneous polarization which is indicated by the black arrows in Fig. 2.

Thin films of LN and BTO exhibit some differences in their domain structures. Firstly, LN is typically produced as a single domain film where the polarization of every unit cell is aligned in the same direction. This is indicated with the single arrow in the plane of the LN film in Fig. 2(a). On the other hand, a raw BTO thin film can have domains with the ferroelectric polarization randomly oriented in one of four 90-degree orientations which is indicated by the double-sided arrows in Fig. 2(b).

The permittivity measurements here are presented differently for each material because of their different domain structures. Fig. 2(c) and (d) show measured permittivity data collected from various literature sources for LN[17–27] and for BTO[28–37], respectively. The permittivity in most LN literature can be measured separately along the $a$-axis and $c$-axis because of the stable single domain structure. The literature values for LN in Fig. 2(c) are therefore categorized by $\varepsilon_a$ (light blue) and $\varepsilon_c$ (dark blue). The permittivity of LN can be described with a constant model in this frequency range. The solid lines in Fig. 2(c) represent the mean literature values of $\varepsilon_a$ = 45 and $\varepsilon_c$ = 27. These values are also used in the analysis of our own measurements to validate our method for measuring the permittivity which is described in further detail in the methods section and Supplementary Note 2. Our work confirms these values as indicated by the black lines.

In contrast to the permittivity measurements of LN, the typical mixed domain structure of BTO makes it difficult to distinguish between $\varepsilon_a$ and $\varepsilon_c$. Instead, the permittivity is often reported without reference to a particular crystal axis. The literature values for BTO (light green) in Fig. 2(d) are therefore the effective permittivity with contributions from both $\varepsilon_a$ and $\varepsilon_c$. Our own thin film measurements are plotted in black.

The BTO permittivity has significant variability among the reported results and cannot be described by a linear fit. From a high-level perspective, the variability is a result of differences in sample quality and crystal growth techniques[31]. Some sources measure ceramics produced from powders[28–30,32,35,37], others measure bulk crystals produced by top-seeded solution growth[30,31], and more recent reports measure thin films grown by metal-organic vapor deposition[33,34], molecular beam epitaxy[36], and chemical solution deposition[38,39]. From a microscopic perspective, sample quality can have significant variation in the quantity of defects, the domain structures and the poling of domains. We discuss how the domain structure and poling affects the permittivity in Supplementary Note 1.

The role of defects is especially relevant because our BTO films are grown by a recently developed epitaxial RF sputtering method[40,41]. This method is suggested to minimize defects since it starts from a stoichiometric BTO target, resulting in a stoichiometric thin film. Other epitaxial deposition methods like molecular beam epitaxy (MBE) or metal-organic chemical vapor deposition (MOCVD) require a careful balance of precursors and still result in crystal defects such as oxygen vacancies that require annealing to be mitigated[42,43]. Defects in the crystal are attributed to a relaxation that falls in the range of 100s of MHz into GHz frequencies[30,31]. In particular, impurities and oxygen vacancies caused by non-stoichiometric crystal growth can drastically affect the strength and the central frequency of this relaxation[30,31]. A second relaxation in the in the range of 50 GHz or higher is associated with the Ti ion hopping between sites along the *a*-axis in the unit cell and is a proposed origin of the large permittivity along the BTO *a*-axis[31,44,45].

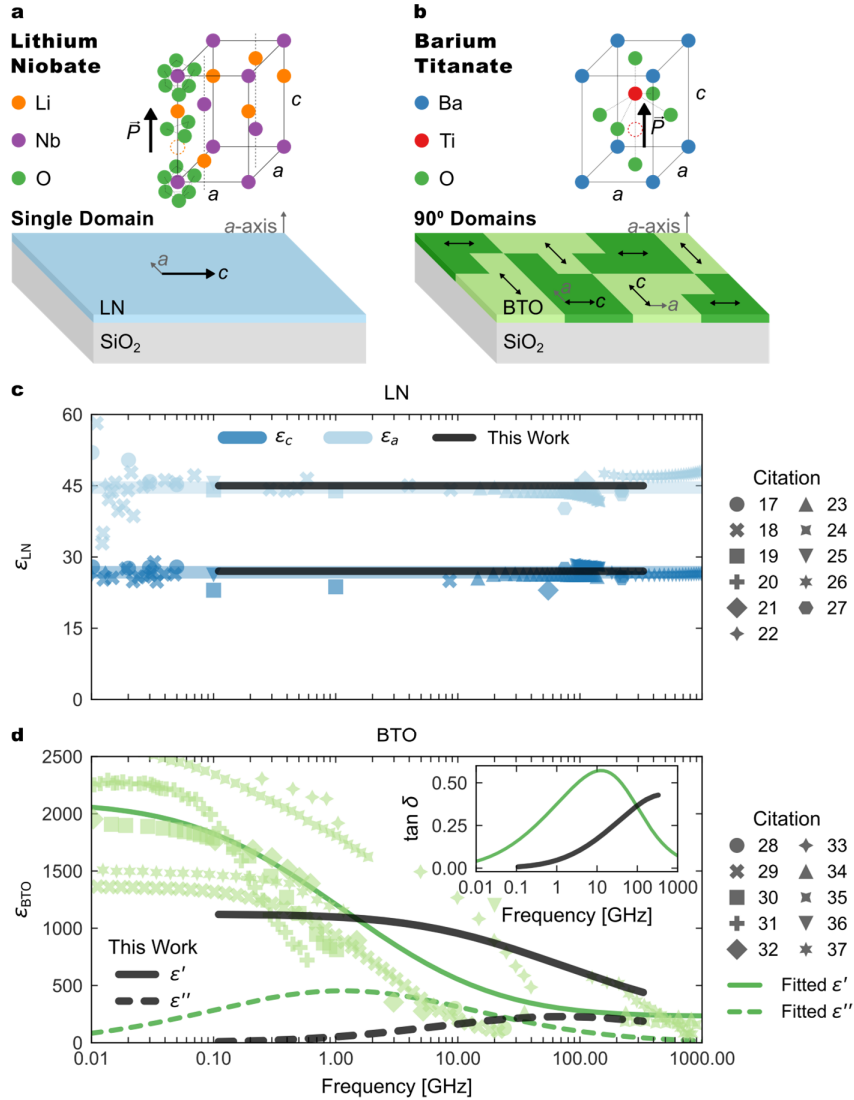

**Fig. 2 | Crystal structure and permittivity of LN and BTO. a-b,** Crystal structures of LN and BTO, respectively, and how they are oriented as thin films. In each unit cell the spontaneous polarization $\vec{P}$ is marked with the black arrow. The "empty" atoms with the dashed outline represent the hypothetical positions of the Li (orange) or Ti (red) ions if the polarization was flipped. In the depiction of the thin films, black arrows are also used to represent the spontaneous polarization. Both have an in-plane *c*-axis. LN has a single domain with only one polarization direction as indicated by the single black arrow. BTO has a multi-domain structure where the *c*-axes of neighbouring domains are oriented at 90° to each other, as indicated by the multiple black arrows and the differently shaded regions. For BTO, the black arrows are double-sided to represent the fact that the raw film might have $\vec{P}$ in either direction. **c,** LN permittivity data along the *a*-axis (light blue) and *c*-axis (dark blue) collected from literature[17–27]. The blue solid lines represent a linear fit to the literature data. The black lines represent the values produced from the model in this work. **d,** BTO permittivity data (light green) collected from literature[28–37]. The green solid and dashed lines are real and imaginary parts, respectively, of the Debye model from equation (2) fitted to the literature data. The permittivity extracted from measurements of the RF-sputtered

BTO of this work are plotted in black. The inset shows the loss tangent $\tan\delta = \varepsilon''/\varepsilon'$ for the literature fit and the data from this work.

We use a Debye model with a normal distribution of logarithmic relaxation frequencies[44] to describe the permittivity of our BTO thin film as well as the collection of literature values. It takes the form of equation (2)

$$\varepsilon'(f) + j\varepsilon''(f) = \varepsilon_\infty + \sum_i \Delta\varepsilon_i \int \mathcal{N}(\gamma, \gamma_{0_i}, \sigma_i) \frac{\gamma}{\gamma - jf} d\gamma$$

$$\mathcal{N}(\gamma, \gamma_{0_i}, \sigma_i) = \frac{1}{\sigma_i \sqrt{2\pi}} e^{-\frac{1}{2}\left(\frac{\log\gamma - \log\gamma_{0_i}}{\sigma_i}\right)^2}.$$

(2)

The summation accounts for the potential to have multiple relaxations as described above (e.g. $i = 1$ for a defect-related relaxation, $i = 2$ for the Ti ion relaxation, $i = 3$ for domain poling, etc.). $\gamma$ is the relaxation frequency, $\Delta\varepsilon_i$ is the relaxation strength and $\varepsilon_\infty$ is the constant permittivity that the model approaches at the high-frequency limit. The normal distribution of logarithmic relaxation frequencies $\mathcal{N}(\gamma, \gamma_{0_i}, \sigma_i)$ is described by its center relaxation frequency $\gamma_{0_i}$ and its width in log-space $\sigma_i$. Table 1 contains the fitted parameters for the data in Fig. 2(d) where the Debye model has been plotted for our measurements (black line) and the collection of literature data (green line). We obtained the best fit for the trend in literature data with only one relaxation. A high frequency relaxation related to the Ti ion could not be fit because the data in literature at these frequencies are too sparse. For our data, the model gives a best fit with only a high frequency relaxation. This result implies that our BTO film has minimal defects as is expected for the RF-sputtered films. Additionally, our BTO is partially poled to reduce antiparallel domains which reduces the permittivity especially at lower frequencies[46].

We also show the imaginary parts of the permittivity $\varepsilon''$ with the dashed lines in the inset of Fig. 2(d). $\varepsilon''$ is related to the dielectric loss in the material.

Since we are only able to measure up to 330 GHz, we fix $\varepsilon_\infty$ to the value from the literature data where measurements from 330 GHz to 1 THz and beyond are available from multiple sources[34,35,37]. The literature data has a much clearer consensus at THz frequencies than in the MHz and GHz range.

Table 1: Fitted parameters of the Debye model in equation (2) for the BTO permittivity data in Fig. 2(d). The uncertainty values are the standard errors of the fits. Note also that $\sigma$ represents the width of the distribution of $\log(\gamma)$. *This value is fixed to the one from literature data because our equipment limits us to 330 GHz. This is supported by multiple sources[34,35,37] that report permittivity up to 1 THz.

| Permittivity Data | $\gamma_0$ [GHz] | $\sigma$ | $\Delta\varepsilon$ | $\varepsilon_\infty$ |
|---|---|---|---|---|
| Literature | 1.15 ± 0.30 | 0.92 ± 0.20 | 1862 ± 149 | 231 ± 79 |
| This Work | 72.6 ± 2.3 | 0.87 ± 0.02 | 905 ± 26 | 231* |

## Pockels Coefficient

Here, we discuss the Pockels effect in LN and BTO. The Pockels tensors for both materials using the reduced Voigt notation are given in equation (3)[47].

$$r_{ij,\,LN} = \begin{bmatrix} 0 & -r_{22} & r_{13} \\ 0 & r_{22} & r_{13} \\ 0 & 0 & r_{33} \\ 0 & r_{42} & 0 \\ r_{42} & 0 & 0 \\ r_{22} & 0 & 0 \end{bmatrix}, \quad r_{ij,\,BTO} = \begin{bmatrix} 0 & 0 & r_{13} \\ 0 & 0 & r_{13} \\ 0 & 0 & r_{33} \\ 0 & r_{42} & 0 \\ r_{42} & 0 & 0 \\ 0 & 0 & 0 \end{bmatrix} \quad (3)$$

The magnitude of the refractive index change depends on the orientation of both the optical ($\vec{E}_{Optical}$) and RF ($\vec{E}_{RF}$) fields relative to the crystal orientation. The effective Pockels coefficient $r_{eff}$ is the sum of the contributions from each elements in the Pockels tensor depending on the orientations of $\vec{E}_{Optical}$ and $\vec{E}_{RF}$ relative to the crystal axes[48,49]. Fig. 3(a) illustrates these orientations in the context of the phase shifters in this work. In this configuration the analytical equation for the effective Pockels coefficient in both LN and BTO takes the following form, which is derived in Supplementary Note 3.

$$r_{eff}(\theta) = \sum_{\varphi} v_{\varphi} (\sin^2(\theta + \varphi) \sin(\theta + \varphi) (r_{13} + 2r_{42}) + r_{33} \sin^3(\theta + \varphi)) \quad (4)$$

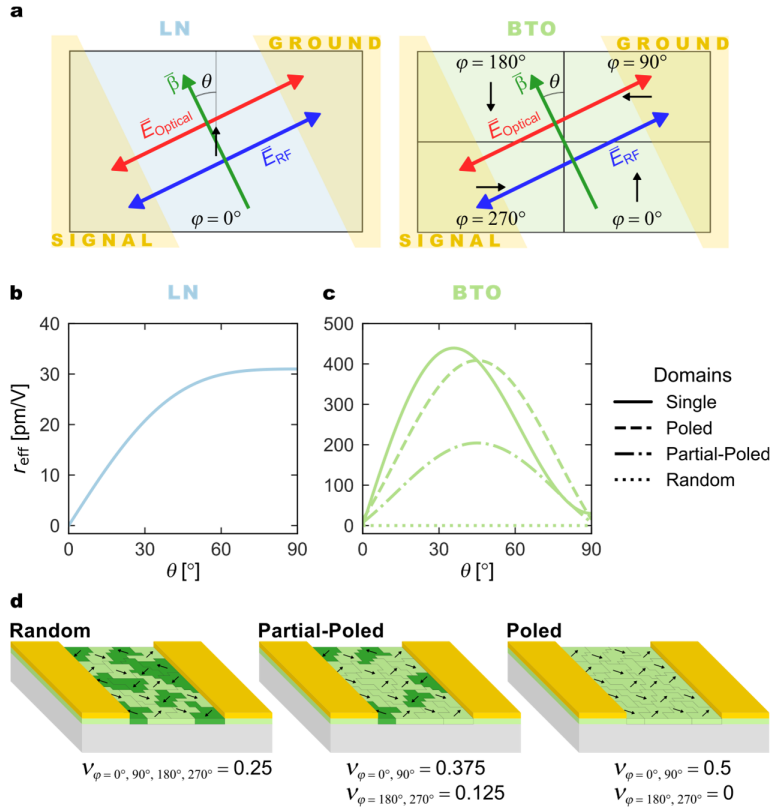

**Fig. 3 | Phase shifter orientation relative to crystal domains and the effective Pockels coefficient. a**, Schematic depicting the phase shifter orientation relative to the domain orientations in the LN and BTO thin films. The orientation of the ferroelectric polarization domains is shown by the black arrows and denoted by the angle $\varphi$. The LN thin film has only a single polarization direction $\varphi = 0°$ while BTO has four possible orientations of $\varphi = 0°$, 90°, 180°, 270°. The areas shaded in gold represent the signal and ground electrodes which also serve as optical waveguides. The phase shifters are oriented such that the propagation direction (green arrow) is at an angle $\theta$ relative to the crystal c-axis. The optical (red arrow) and RF (blue arrow) electric fields are both polarized perpendicular to the propagation direction. **b-c**, Analytically calculated effective Pockels coefficients according

to equation (4) for LN and BTO, respectively. The solid line shows the case where all domains are oriented in the same direction ($v_{\varphi=0°}$ = 1). The dashed and dotted lines show the case for films with domains oriented at 90 degrees to each other where domains are fully poled (dashed), partially poled (dot-dashed) or randomly distributed (dotted). **d** Illustrations of the domain polarizations in BTO for the three cases of poling along with their domain fractions $v_\varphi$. Domains with the preferred polarization direction are coloured in light green while the anti-parallel domains are coloured in dark green.

Following the convention in a previous analysis[43], here we have also introduced the terms $\varphi$ and $v_\varphi$ to account for the domain orientation and the fraction of domains with a particular orientation, respectively. $\varphi$ can take one of four possible values (0°, 90°, 180°, 270°) representing the direction of the ferroelectric polarization in the plane of the thin film. $v_\varphi$ is the relative fraction of domains in the thin film with orientation $\varphi$ and can take a value between 0 and 1 with $\sum_\varphi v_\varphi$ = 1. To demonstrate the effects of different domain distributions we compare $r_{\text{eff}}$ as given by equation (4) in Fig. 3(b) for LN and Fig. 3(c) for four cases in BTO. The curves were calculated using literature values for the clamped Pockels coefficients measured near 10 MHz for LN[25] and BTO[50]. The single domain case (solid line, $v_{\varphi=0°}$ = 1) applies to the LN films but is also plotted for a hypothetical single domain BTO film. For LN $r_{\text{eff}}$ is maximal when $\vec{E}_{\text{Optical}}$ and $\vec{E}_{\text{RF}}$ are parallel to the *c*-axis ($\theta$ = 90°) because $r_{33}$ is the largest Pockels coefficient in LN. The multi-domain structure of BTO thin films results in a maximum $r_{\text{eff}}$ when $\theta$=45°. This is shown for the three cases where a film is poled (dashed line), partial-poled (dash-dot) or random (dotted). Fig. 3(d) illustrates these three cases where the black arrows represent the polarization direction and the light green domains are polarized in the preferred direction. The maximum value of $r_{\text{eff}}$ is achieved for the poled case when all ferroelectric polarizations in the domains are aligned in a preferred direction. When domains have polarizations in opposite directions the net Pockels shift will be zero, as is the case for the random film. The frequency dependence remains approximately the same regardless of the domain distribution because it originates from the ions in the BTO unit cell. Polarizations can be aligned preferentially using an external DC bias, however, in our measurements this was not possible at the highest frequencies due to the limitations of our instruments. The partial-poled case is the most realistic case for the BTO films in our measurements with domain fractions that were reported for similar thin films[43]. This state is reached after poling the film with a DC bias and then removing the bias before performing EO measurements. Further discussions of this poling procedure and of the domain fractions are given in Supplementary Note 3 and the methods section, respectively.

Effective Pockels coefficients of the LN and BTO films were extracted from phase modulation measurements using the measurement setup and analysis procedure described in the methods section. $r_{\text{eff}}$ was measured for different phase shifters oriented at varying angles $\theta$ to the substrate. The set of $r_{\text{eff}}(\theta)$ data points at each frequency was to fit equation (4) with domain fractions corresponding to the partial-poled case. The fitting process yields values for $r_{13} + 2r_{42}$ and $r_{33}$ as explained in further detail in Supplementary Note 3. Fig. 4 shows the results of this measurement and fitting process for both LN (blue) and BTO (green) across all measured frequencies. In both cases, the error bars represent ± two standard errors of the fit.

The LN Pockels coefficients are presented in Fig. 4(a). LN is a good reference material because its EO properties are well-known and it is not expected to show any dispersion in the measured frequency range. Indeed, the data for both $r_{33}$ (dark blue) and $r_{13} + 2r_{42}$

(light blue) are flat over three orders of magnitude in frequency. The solid lines are linear fits to the measured data with $r_{33}$ = 26.9 pm/V and ½$(r_{13} + 2r_{42})$ = 15.0 pm/V. We note that the lower values compared to literature[25] are likely due to the fact that we have used an optical wavelength of 1550 nm instead of 633 nm, which has been shown to influence the Pockels coefficient in ferroelectrics[50].

BTO's Pockels coefficients are presented in Fig. 4(b). The $r_{33}$ data (dark green) shows a dispersion step at a few hundred MHz while the $r_{13} + 2r_{42}$ data (light green) shows strong dispersion above 10 GHz. The solid lines represent the same type of Debye model that was used to model the BTO permittivity. For $r_{33}$ the data is fitted with parameters $\gamma_0$ = 565 MHz, $\sigma$ = 0.06, $S_R$ = 75 pm/V and $r_\infty$ = 60 pm/V (analogous to $\varepsilon_\infty$). The $r_{33}$ value of 125 pm/V at 100 MHz is in close agreement with recent measurements of similar films for frequencies below 10 MHz[41]. In the methods section we show that the Pockels coefficient can be expressed as a function of the RF permittivity through Miller's rule[51]

$$r \approx 2\delta \frac{(n_0^2 - 1)^2}{n_0^4}(\varepsilon - 1) , \qquad (5)$$

where $n_0$ = 2.26 is the optical refractive index for BTO at 1550 nm and $\delta$ is Miller's coefficient. The $r_{13} + 2r_{42}$ model in Fig. 4(b) is from equation (5) where $\varepsilon$ is calculated from the coefficients in Table 1. The best fit was found to be $\delta$ = 0.327 ± 0.002 pm/V. This gives values for ½$(r_{13} + 2r_{42})$ of 481 pm/V at 100 MHz and 191 pm/V at 330 GHz. equation (5) is in excellent agreement with the measurements and lends credence to the empirical Miller's rule. Much like the permittivity, BTO's $r_{42}$ coefficient is remarkably large at lower frequencies but decreases substantially across the measured frequency range. Given the link between first and second order susceptibilities, the measured Pockels coefficients imply that the permittivity dispersion is related to the same underlying phenomenon as the $r_{13} + 2r_{42}$ dispersion. In fact, these results suggest that the decrease in the overall permittivity of the film can be attributed entirely to a decrease in $\varepsilon_a$. Since the frequency response of the $r_{33}$ coefficient is flat above 1 GHz, it would be reasonable to assume that the permittivity along the BTO c-axis is likely constant across this frequency range too.

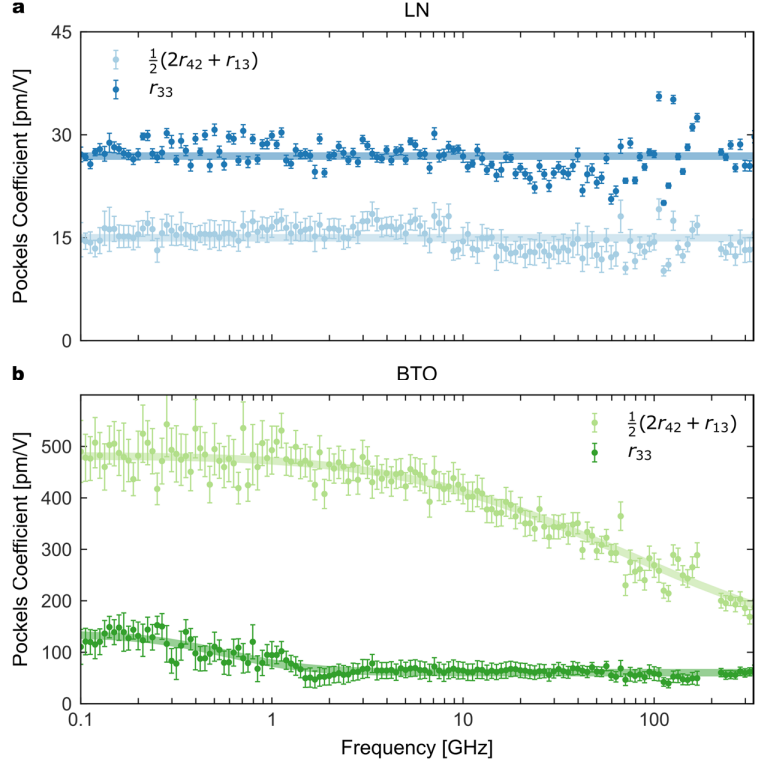

**Fig. 4 | Pockels coefficients of LN and BTO. a**, Pockels coefficient measurements for LN with $r_{33}$ in dark blue, $r_{42}$ in light blue and linear fits to the data in solid lines. The fitted values are $r_{33}$ = 26.9pm/V and ½$(r_{13} + 2r_{42})$ = 15.0 pm/V. **b**, Pockels coefficient measurements for BTO with $r_{33}$ in dark green, $r_{42}$ in light green and model fits to the data in solid lines. The $r_{33}$ data is fit to a Debye model with $\gamma_0$ = 565 MHz, $\sigma$ = 0.06, $S_R$ = 75 pm/V and $r_\infty$ = 60 pm/V. The ½$(r_{13} + 2r_{42})$ is fit with equation (5) where $\varepsilon$ is calculated from the Debye model of equation (2) using the same coefficients as the measured permittivity in Table 1. The error bars in both plots represent two standard errors of the fit, relating to the fitting of measured $r_{\text{eff}}$ data to equation (4). Note that typically $2r_{42} \gg r_{13}$ so that $r_{13} + 2r_{42} \approx 2r_{42}$. We therefore plot this parameter as ½$(r_{13} + 2r_{42}) \approx r_{42}$.

## The Frequency Dependence of the Pockels Shift in BTO Devices

The results above provoke the following question: if BTO's $r_{42}$ decreases with frequency, is it unfavourable to BTO's suitability for high-speed devices? In this section we show that the answer depends on the device geometry by comparing BTO in photonic and plasmonic structures.

To start this discussion we introduce the simplistic, hypothetical structure in Fig. 5(a). Two metal electrodes with a potential difference $V$ are placed on each side of a BTO layer with width $w_{\text{BTO}}$ and RF permittivity $\varepsilon_{\text{BTO}} = 1 + \chi_{\text{BTO}}^{(1)}$. The electrodes are separated from the BTO layer by a cladding material with width $w_C$ and RF permittivity $\varepsilon_C = 1 + \chi_C^{(1)}$. The structure can mimic either a photonic modulator[43] when $w_C \gtrsim w_{\text{BTO}}$ or a plasmonic modulator[49] when $w_C = 0$. The refractive index change $\Delta n_{\text{BTO}}$ as a function of frequency in this structure is given by equation (6), which is derived in Supplementary Note 4

$$\Delta n_{\text{BTO}}(f) = \eta \frac{(1 + R_w)(\varepsilon_{\text{BTO}}(f) - 1)}{1 + R_w \frac{\varepsilon_{\text{BTO}}(f)}{\varepsilon_C}}. \tag{6}$$

Here, $R_w = 2w_C/w_{\text{BTO}}$ is the ratio of total cladding width to BTO width and the factor $\eta$ is a group of constant terms that are irrelevant to the present analysis. To arrive at equation (6) we exploited the boundary condition $\varepsilon_C E_C = \varepsilon_{BTO} E_{BTO}$ for electric fields that are perpendicular to the cladding-BTO interface (Fig. 5(a), blue line). If $\varepsilon_{BTO}$ decreases then $E_{BTO}$ must increase to maintain continuity at the interface. As a result, a decreasing $\varepsilon_{BTO}$ can counteract a decreasing $r_{BTO}$ (see equation (1)). We have also used Millers rule – which relates $\chi^{(1)}_{BTO}$ and $\chi^{(2)}_{BTO}$ – to eliminate any dependence on $\chi^{(2)}_{BTO}$ or $r_{BTO}$.

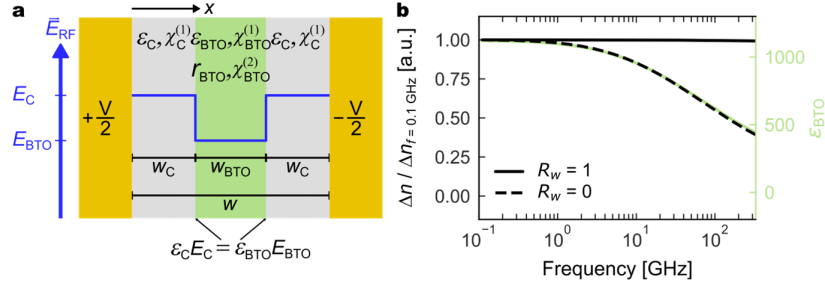

**Fig. 5 | Frequency dependence of the Pockels shift in a device. a**, Schematic of a hypothetical device with a potential difference $V$ across two electrodes separated by a width $w$, a cladding material (gray) with permittivity $\varepsilon_C$, susceptibility $\chi^{(1)}_C$ and width $w_C$, and an electro-optic material (green) with permittivity $\varepsilon_{BTO}$, susceptibility $\chi^{(1)}_{BTO}$ and width $w_{BTO}$. The blue line marks how the boundary condition at the bottom of the diagram affects the electric field strength in each material assuming $\varepsilon_{BTO} > \varepsilon_C$. **b**, Refractive index change in the BTO material according to equation (6) as a function of frequency for varying ratios $R_w$ with $\varepsilon_C$ = 4. The measured permittivity of BTO in this work (see Table 1) was used as the input for $\varepsilon_{BTO}$ and is plotted in green. $\Delta n$ is normalized to its low frequency value at 100 MHz to highlight the frequency dependence for each case.

There are two interesting cases of $R_w$ corresponding to plasmonic and photonic devices. Firstly, in the plasmonic devices where $w_C$ and $R_w$ are zero, equation (6) simplifies to

$$\Delta n_{BTO}(R_w = 0) = \eta(\varepsilon_{BTO} - 1) \ . \qquad (7)$$

This implies that any frequency dependence of $\varepsilon_{BTO}$ will also be apparent in $\Delta n_{BTO}$. Fig. 5(b) shows the frequency dependence of $\Delta n_{BTO}$ for various $R_w$ values and with $\varepsilon_C$ = 4. For $R_w$ = 0 (dashed line) the shape of the curve closely follows the frequency response of $\varepsilon_{BTO}$ (green) that was measured in this work. Secondly, typical photonic devices will have $R_w \approx 1$ or larger. For the case of large $R_w$, equation (6) can be approximated by

$$\Delta n_{BTO}(R_w \gg 1) \approx \eta \varepsilon_C \frac{\varepsilon_{BTO} - 1}{\varepsilon_{BTO}}, \qquad (8)$$

where the last term is approximately unity when $\varepsilon_{BTO}$ is large. The significance of this cannot be understated; $\Delta n_{BTO}$ essentially does not depend on $\varepsilon_{BTO}$ or $r_{BTO}$ when $\varepsilon_{BTO}$ is sufficiently large. Fig. 5(b) shows that this is true even for $R_w$ = 1 (solid line) where $\Delta n$ is constant.

We note that the focus here is on the frequency response which is why $\Delta n_{BTO}$ is plotted relative to its low-frequency value in Fig. 5(b). Small values of $R_w$ still give large values of $\Delta n_{BTO}$ in absolute terms, which is expected from plasmonic devices. However, this analysis highlights a tradeoff for BTO devices between the highly efficient response of plasmonics and the frequency-independent response of photonics. Nevertheless, it might explain why previous demonstrations of BTO plasmonic modulators show a frequency dependent EO response while the photonic counterparts do not[43,49,52].

## Conclusion

The frequency response of the permittivity and Pockels coefficients of LN and BTO have been measured. The data cover a broad frequency range important to a variety of developing technologies including high-speed communications, quantum networks, programmable photonics and reconfigurable metasurfaces. It provides a foundation for the design of novel devices operating up to the highest frequencies. The phase shifter and measurement method provide an avenue for exploring the EO properties of new materials as they are developed. Finally, the insights into the relationships between the refractive index change, the BTO permittivity, and the geometry of the device in which it is used may spark innovative ideas for overcoming the frequency dependence of EO properties.

## Acknowledgements

We wish to thank Agham Posadas and Alexander Demkov of La Luce Cristallina for providing us with thin film BTO. We also wish to thank the Cleanroom Operations Team of the Binnig and Rohrer Nanotechnology Center (BRNC) for their help and support. This work was funded in part by the EC H2020 projects NEBULA (871658), PlasmoniAC (871391), and the SNF project (208094).

## Author Contributions Statement

DC: conceptualization, theory, simulation, device design, fabrication, RF and EO characterization, data analysis, methodology, visualization, writing

MK: fabrication, EO characterization, methodology, writing

JW & DM: theory, fabrication, methodology

AM: theory, fabrication, EO characterization, methodology

YF: fabrication, writing, supervision

ME, YL & HW: RF characterization, data analysis, supervision

JL: conceptualization, writing, supervision

## Competing Interests Statement

The authors declare no competing interests.

# Methods

### Device Fabrication

The phase shifters were fabricated on BTO-on-insulator and LN-on-insulator substrates with a 485 nm BTO layer or a 300 nm LN layer and a 2 µm buried oxide layer. Access waveguides were patterned with electron beam lithography (EBL) and transferred into the BTO/LN layers with ICP-RIE. The samples were annealed in an $O_2$ atmosphere at 550°C for 1 hour to reduce optical propagation losses[42]. The 100 nm $SiO_2$ spacer layer was deposited with plasma-enhanced chemical vapor deposition (PECVD). Next, the grating couplers were fabricated from 130 nm of amorphous silicon that was deposited with PECVD, patterned with EBL and etched with ICP-RIE using HBr plasma. The electrodes were deposited with a liftoff process where EBL was used for patterning and electron beam evaporation was used to deposit 200 nm of gold. In a final step, the PMMA cladding was spin-coated on to the sample.

This simple process has multiple benefits. The dimensions of the structure are well-defined and easy to reproduce across fabrication runs. The layer thicknesses are precisely controlled by modern deposition techniques and can be confirmed through ellipsometry. In Supplementary Note 5 we show through simulations that uncertainties in layer thickness have only a small effect on the phase shifter's voltage-length product. The only lithographic patterning required in the phase shifter section is for the electrodes. There is no critical alignment with other lithography layers and the µm-sized gap between the electrodes is easily achieved with photolithography. Additionally, there is no etching required in the phase shifter section. This means that there is little uncertainty with respect to etching depths, waveguide widths or sidewall angles. Furthermore, there is no etching byproduct or physical damage introduced to the film that may alter the EO properties.

### EO Measurement Setup

The setup used for electro-optic measurements is described here. A tunable laser source (TLS) provides the optical carrier which passes through a polarization controller (PC) before being coupled to the chip (PIC) with grating couplers (GC). Waveguides route the optical signal to the phase modulators which are driven by an external source ($f_{RF}$) through RF probes. The phase shifters are driven as an open load at the end of the probe without a termination because they have a length of only 25 µm for BTO or 50 µm for LN. All measurements were performed without a DC bias, but after the poling process described in Supplementary Note 3. The modulated optical signal is coupled out from the chip with another grating coupler and its spectrum is recorded with an optical spectrum analyzer (OSA). Four different sources were used to generate modulating signals across the full frequency range. An analog signal generator (Agilent E8257D) was used to generate signals up to 70 GHz. The same signal generator was used to drive various frequency multipliers to generate frequencies above 70 GHz. The range between 70-110 GHz was generated with a frequency multiplier (Radiometer Physics AFM6 75-110 +10). VNA extension modules (VDI VNAX WR6.5 and WR3.4) were used as frequency multipliers for the ranges between 110-170 GHz and 220-330 GHz.

VNA measurements were performed using a Keysight PNA-X N5247B in a 1-port configuration and using the various VNAX extension modules from VDI to reach up to 330 GHz.

### Extracting Permittivity from VNA Measurements

Our method for measuring the permittivity is based on $S_{11}$ reflection measurements performed with a vector network analyzer (VNA). The $S_{11}$ coefficient is equivalent to the reflection coefficient and is determined by $S_{11} = (Z - Z_0)/(Z + Z_0)$ where $Z$ is the phase shifter impedance and $Z_0$ = 50 Ω is the reference impedance of the system. The total impedance of a phase shifter can therefore be determined from

$$Z = Z_0 \cdot \frac{1 + S_{11}}{1 - S_{11}}. \tag{9}$$

The measured $Z$ is then compared to a theoretically calculated impedance $Z_{\text{theory}}(f, \varepsilon_{\text{EO}})$ of an equivalent circuit model for the phase shifter under test where $f$ is the frequency and $\varepsilon_{\text{EO}}$ is the complex permittivity of the EO layer. The measured permittivity is determined by finding the $\varepsilon_{\text{EO}}$ that makes the best match between $Z_{\text{theory}}$ and the measured $Z$. More details on the equivalent circuit and the calculation of its impedance elements can be found in Supplementary Note 2. To validate our model, we use the average literature values for the LN permittivity. The model depends primarily on the in-plane permittivity so $\varepsilon_{\text{EO}} = \varepsilon_c$ = 27 is inserted into the calculation of $Z_{\text{theory}}(f, \varepsilon_{\text{EO}})$. The model gives an excellent fit to the measured impedance, see Supplementary Note 2. The BTO impedance is not well fit for any constant value of $\varepsilon_{\text{EO}}$. Instead, we use equation (2) to calculate $\varepsilon_{\text{EO}}$ and the impedance model $Z_{\text{theory}}(f, \varepsilon_{\text{EO}}(\gamma_0, \sigma, S_R, \varepsilon_\infty))$ is dependent on the same parameters as the Debye permittivity model of equation (2). The complex permittivity given by the Debye model is essential to the impedance model for the purpose of calculating the dielectric loss. The results for the measured and fitted impedance of BTO can also be found in Supplementary Note 2.

### EO Simulations

This section describes the procedure for calculating theoretical values of the voltage-length product $V_\pi L$. The half-wave voltage $V_\pi$ is defined as the voltage required to change the phase of an optical signal by $\pi$. If $\Delta\beta = \beta|_{1\text{ V}} - \beta|_{0\text{ V}}$ is the difference between the propagation constant of an optical mode with and without an applied voltage, then the total phase difference after propagation through a modulator with length $L$ is

$$\Delta\varphi = \Delta\beta L = \frac{2\pi L}{\lambda}\Delta n_{\text{eff}} = \frac{2\pi L}{\lambda}\left(n_{\text{eff}}|_{1\text{ V}} - n_{\text{eff}}|_{0\text{ V}}\right). \tag{10}$$

To calculate the effective indices $n_{\text{eff}}|_{1\text{ V}}$ and $n_{\text{eff}}|_{0\text{ V}}$ electro-optic simulations were performed with Lumerical's MODE and DEVICE simulation software. In a first step, electrical simulations were performed in DEVICE to calculate the RF electric field profile when a 1 V potential difference is applied to the electrodes. A small-signal AC perturbation was used to calculate the field profile at different RF frequencies. Electrical simulations were performed for various permittivities of the LN/BTO layers. Next, the simulated RF electric field was used with the full Pockels tensor for each material to calculate the spatially varying refractive index profile in the LN/BTO layer. Values for the Pockels coefficients were taken from literature[25,53]. Optical simulations were performed in MODE to find the effective index $n_{eff}$ of the mode supported by the phase shifter. Two simulations were performed: one without the external RF field to get $n_{eff}|_{0\text{ V}}$, and one with the external RF field to get $n_{eff}|_{1\text{ V}}$. To achieve a phase shift $\Delta\varphi = \pi$ with the $\Delta n_{eff}$ calculated from an applied voltage of 1 V, the phase shifter must have a length $L_\pi$ equal to

$$L_\pi = (\Delta\varphi = \pi)\frac{\lambda}{2\pi \Delta n_{eff}} = \frac{1}{2}\frac{\lambda}{\Delta n_{eff}}. \tag{11}$$

As implied by its name, the voltage-length product is the product of the applied voltage and the phase shifter length. Since the applied voltage in this case is 1 V the voltage-length product is then $V_\pi L = V L_\pi = (1)L_\pi = L_\pi$. The simulated voltage-length product is therefore given by equation (11) which is restated as

$$V_\pi L = \frac{1}{2}\frac{\lambda}{\Delta n_{eff}}. \tag{12}$$

### Extracting $r_{eff}$ from Phase Modulation Measurements

EO modulators are typically employed to change the phase of an optical mode using an electrical drive signal. Such phase changes are also accompanied by changes to the optical spectrum. New frequencies are generated at $\omega_{\pm k} = \omega_0 \pm k\omega_{RF}$ where $\omega_0$ is the optical carrier frequency, $\omega_{RF}$ is the RF modulation frequency and k = ±1, ±2, … is an integer. The intensities of each frequency in the optical spectrum at the output of the modulator are given by Bessel functions and depend on the modulation efficiency[54]. Crucially, this allows the EO strength to be extracted from optical intensity measurements rather than phase change measurements which can be troublesome, especially at high frequencies.

The modulation efficiency, or alternatively the half-wave voltage $V_\pi$, of a phase shifter can be calculated from the intensity ratio between the optical carrier and the first modulation sideband as described by equation (13)[54]

$$V_\pi = V_{RF}\frac{\pi}{2}\sqrt{\frac{I(\omega_0)}{I(\omega_0 \pm \omega_{RF})}}, \tag{13}$$

where $I(\omega_0)$ is the intensity of the optical carrier, $I(\omega_0 \pm \omega_{RF})$ is the intensity of the first modulation sideband and $V_{RF}$ is the peak voltage of the electrical modulation signal across the phase shifter electrodes. The two intensities can be measured with an optical spectrum analyzer, thus, if one knows the $V_{RF}$ then $V_\pi$ can be easily calculated.

Determining $V_{RF}$, however, is not always trivial because it depends on the impedance and reflection characteristics of the phase shifter. Conveniently, the frequency-dependent reflection is already available from the electrical $S_{11}$ measurements from which the permittivity was extracted. The phase shifter can be conceptualized as a termination at the end of an RF transmission line that delivers the driving signal. In this case, the voltage across the electrodes of the phase shifter is equal to $V_{RF} = V_{drive}(1 + S_{11})$. In other words, $V_{RF}$ is the sum of the incoming and reflected electrical waves which have amplitudes $V_{drive}$ and $S_{11} \cdot V_{drive}$, respectively. $V_{RF}$ can therefore be determined with high accuracy by combining the measurements of $S_{11}$ along with a calibration of the RF losses between the RF source and the phase shifter.

Next, an expression is needed to relate $V_\pi$ to $r_{\text{eff}}$. We start from the basis of equation (1) in the main text and equation (10) from the previous section. To relate the two equations requires an extra term that describes how much the effective index of the optical mode will change ($\Delta n_{\text{eff}}$) for a given change in the refractive index of the EO material ($\Delta n_{\text{EO}}$). This term is known as the field interactor factor $\Gamma$[15]

$$\Gamma = \frac{\Delta n_{\text{eff}}}{\Delta n_{\text{EO}}} . \tag{14}$$

$\Gamma$ accounts for the spatial variation of the RF and optical fields as well as the overlap between the two fields and the active material (i.e. LN or BTO). Inserting $\Delta n_{\text{eff}} = \Delta n_{\text{EO}} \Gamma$ into equation (10) and then using equation (1) to replace $\Delta n_{\text{EO}}$ gives

$$\Delta \varphi = \frac{\pi L}{\lambda} n_{\text{EO}}^3 r_{\text{eff}} \Gamma \frac{V}{w_{\text{gap}}} , \tag{15}$$

where the electric field $E$ in equation (1) has been replaced by $V/w_{\text{gap}}$ with $w_{\text{gap}}$ being the width of the gap between the electrodes. Once again setting $\Delta \varphi = \pi$ as per the definition of $V_\pi$ gives an expression for the voltage-length product

$$V_\pi L = \frac{\lambda w_{\text{gap}}}{n_{\text{EO}}^3 r_{\text{eff}} \Gamma} . \tag{16}$$

The effective Pockels coefficient $r_{\text{eff}}$ can then be calculated from the measured $V_\pi$ after rearranging equation (16) to be

$$r_{\text{eff}} = \frac{\lambda w_{\text{gap}}}{n_{\text{EO}}^3 \Gamma V_\pi L} . \tag{17}$$

## Distribution of Ferroelectric Domains in BTO

This section describes in further detail the framework for describing $r_{\text{eff}}$ in a film with multiple ferroelectric domains. First, consider the case where the thin film consists entirely of a single domain ($\varphi = 0°$, $v_{\varphi=0°} = 1$) as is the case for LN. For materials like LN where the $r_{33} = r_{zzz}$ Pockels coefficient is the largest, $r_{\text{eff}}$ as calculated by equation (4) is maximal for $\theta = 90°$ when the optical and RF fields are polarized along the c-axis (i.e. z-axis). Fig. 3(b) shows the dependence of $r_{\text{eff}}$ on $\theta$ for LN using values[25] of $r_{33}$ = 31 pm/V, $r_{42}$ = 18 pm/V and $r_{13}$ = 9 pm/V. In contrast, BTO's largest Pockels coefficient is the $r_{42}$ so it reaches a maximal $r_{\text{eff}}$ for some angle $\theta$ that is not parallel to one of the crystal's primary axes. This happens because $r_{42} = r_{yzy}$ requires $\vec{E}_{\text{Optical}}$ and $\vec{E}_{\text{RF}}$ components along two orthogonal crystal axes. The phase shifters in this work, however, have $\vec{E}_{\text{Optical}}$ and $\vec{E}_{\text{RF}}$ aligned in the same direction. The device and the electric fields must therefore be rotated by an angle $\theta$ such that there will be projections of the fields onto two orthogonal crystal axes. The dependence of $r_{\text{eff}}$ versus $\theta$ for a hypothetical single domain BTO thin film is shown by the solid curve in Fig. 3(c). The maximum $r_{\text{eff}}$ in this case is for waveguides aligned with $\theta \approx 36°$. For BTO we used values[50] of $r_{33}$ = 30 pm/V, $r_{42}$ = 560 pm/V and $r_{13}$ = 6 pm/V. Both the LN and BTO values used in the analytical calculations are the clamped Pockels coefficients measured at frequencies near 10 MHz.

BTO thin films with an in-plane c-axis typically consist of multiple domains oriented at right angles to each other. If the domains are randomly distributed and all orientations of $\varphi$ are equally likely ($v_{\varphi=0°,90°,180°,270°} = 0.25$) then the Pockels effect from neighbouring domains will cancel out on average and $r_{\text{eff}}$ is zero for all $\theta$ as shown by the dotted line in

Fig. 3(c). A non-zero $r_{\text{eff}}$ requires an unequal distribution of domain orientations. This can be achieved through the application of a DC bias field between the two electrodes, which can flip the ferroelectric polarization in a domain by 180°. For example, a bias applied between the ground and signal electrodes in Fig. 3(a) could flip the $\varphi$ = 180° domain to a $\varphi$ = 0° domain and the $\varphi$ = 270° domain to a $\varphi$ = 90° domain. This effect is illustrated in Fig. 3(d) where the arrows represent the polarization direction within each domain. In a multi-domain BTO film the ideal distribution of domains should have no anti-parallel domains in order to maximize $r_{\text{eff}}$. This means the relative domain fractions for a fully poled film should be $\nu_{\varphi=0°,90°}$ = 0.5 and $\nu_{\varphi=180°,270°}$ = 0 or vice versa. The dashed line in Fig. 3(c) shows that $r_{\text{eff}}$ in this situation is maximal when $\theta$ = 45° and it is slightly less than the maximum for a single-domain film.

A more realistic case for our experiments consists of a majority domains polarized in a preferred direction, but with some anti-parallel domains still present. For this partially poled case we use domain fractions of $\nu_{\varphi=0°,90°}$ = 0.375 and $\nu_{\varphi=180°,270°}$ = 0.125. These values were reported for similar thin films without the presence of an external bias field but after applying an initial poling step to set the majority of the domains in the correct orientation[43]. The $r_{\text{eff}}$ for this case is plotted in the dash-dot line in Fig. 3(c) and is also maximal for $\theta$ = 45°. In fact, the curve for this partial-poled case is the same as that for the fully poled case but scaled by a constant. This is a result of the counteracting contributions from the anti-parallel domains to the overall EO effect.

### Relating $\chi^{(1)}$ and $\chi^{(2)}$ Through Miller's Rule

The measured permittivity values give an insight to the frequency response of the Pockels coefficients through Miller's rule. Miller's rule states that there is a nearly constant ratio $\delta$ between the 2$^{\text{nd}}$ order susceptibility $\chi^{(2)}(\omega_1, \omega_2, \omega_3)$ that enables a nonlinear process and the product of the 1$^{\text{st}}$ order susceptibilities $\chi^{(1)}(\omega_1)$, $\chi^{(1)}(\omega_2)$, $\chi^{(1)}(\omega_3)$ corresponding to each of the frequencies involved in the process[47,51]. The Pockels effect in an optical phase shifter is essentially a sum/difference frequency process with an optical frequency $\omega_0$, a modulating frequency $\omega_m$ and sum/difference frequencies $\omega_0 \pm \omega_m$. In the form of Miller's rule this looks like equation (18).

$$\delta = \frac{\chi^{(2)}(\omega_0 \pm \omega_m, \omega_0, \omega_m)}{\chi^{(1)}(\omega_0 \pm \omega_m)\chi^{(1)}(\omega_0)\chi^{(1)}(\omega_m)} \tag{18}$$

For the case of the measurements presented here, the optical carrier has a fixed frequency $\omega_0 \approx$ 193 THz ($\lambda$ = 1550 nm) and therefore $\chi^{(1)}(\omega_0)$ is constant. Additionally, $\omega_0$ is much larger than the modulating frequencies between 100 MHz and 330 GHz. This means that $\omega_0 \pm \omega_m \approx \omega_0$ for all $\omega_m$ and therefore $\chi^{(1)}(\omega_0 \pm \omega_m)$ is also constant. If $\delta$, $\chi^{(1)}(\omega_0)$ and $\chi^{(1)}(\omega_0 \pm \omega_m)$ are all constant, then any change in the value of $\chi^{(1)}(\omega_m)$ must also be accompanied by a similarly proportioned change in $\chi^{(2)}(\omega_0 \pm \omega_m, \omega_0, \omega_m)$ and vice versa. Since the dielectric constant can be expressed as $\varepsilon = 1 + \chi^{(1)}$ and the Pockels coefficient can be expressed as $r = 2\chi^{(2)}/n^4$, it can therefore be inferred that when measurements show changes in the dielectric constant as a function of frequency there are similar changes occurring in the Pockels coefficients. This also allows the Pockels coefficient to be expressed as a product of first order susceptibilities

$$r \approx \frac{2\delta}{n_0^4} \left(\chi^{(1)}(\omega_0)\right)^2 \chi^{(1)}(\omega_{\text{m}}) . \tag{19}$$

This can also be expressed in terms of optical refractive index and RF permittivity by considering that $n^2 = \varepsilon = 1 + \chi^{(1)}$. The form of equation (5) in the main text is reached after making the relevant substitutions for the optical and RF susceptibilities

$$r \approx 2\delta \frac{(n_0^2 - 1)^2}{n^4} (\varepsilon - 1) . \tag{20}$$

## Data Availability

The data that support this work are presented in the main text and the Supplementary information. Further data are available from the corresponding authors upon reasonable request.

## Methods-only References

# Barium Titanate and Lithium Niobate Permittivity and Pockels Coefficients from MHz to Sub-THz Frequencies: Supplementary Information

Daniel Chelladurai[1], Manuel Kohli[1], Joel Winiger[1], David Moor[1], Andreas Messner[1,†], Yuriy Fedoryshyn[1], Mohammed Eleraky[2], Yuqi Liu[2], Hua Wang[2] and Juerg Leuthold[1]

[1]Institute of Electromagnetic Fields (IEF), ETH Zurich, Gloriastrasse 35, 8092 Zurich, Switzerland
[2]Integrated Devices, Electronics, And Systems (IDEAS) Group, ETH Zurich, Gloriastrasse 35, 8092 Zurich, Switzerland
[†]Now with: Zurich Instruments AG, 8005 Zurich, Switzerland

## Table of Contents

# 1 Origins of the Frequency Dependence of the Permittivity and Pockels Coefficients

The permittivity and Pockels coefficients are frequency dependent because they originate from piezoelectric, ionic and electronic contributions that die out above their respective excitation frequencies[1,2]. At frequencies from DC to ~1 MHz there are contributions from the mechanical deformation of a lattice due to the piezoelectric effect and from lattice vibrations related to acoustic phonons. A distinction is typically made between the frequency regimes above and below the acoustic phonon frequencies which are typically between 1 and 100 MHz. Below 1 MHz a crystal is said to be free because it is free to deform under the influence of an external electric field. Above ~100 MHz a crystal is said to be clamped because these mechanical deformations no longer occur due to their slow timescale. Nonlinearities related to optical phonons also contribute to the Pockels effect. This is commonly referred to as the ionic contribution and it dies out at optical phonon frequencies which are usually around a few THz. Finally, the highest frequency contributions to EO effects originate from electronic resonances which have characteristic frequencies in the PHz regime. A result of the various contributions is that different materials may have different EO responses in certain frequency ranges. Organic polymers are an example of materials that derive their nonlinearity almost exclusively from electrons[2]. They maintain strong EO effects up to optical frequencies and have been demonstrated in photonic integrated circuits with phase shifters operating up to at least 500 GHz[3] and in antennas operating up to 2.4 THz[4]. However, inorganic materials are often preferred for their temperature stability and compatibility with common fabrication processes.

The primary contribution to the nonlinearities in ferroelectric crystals comes from ionic resonances. One consequence of the EO strength being dominated by acoustic and optical lattice vibrations is that their contributions can be very different between materials. For example, the largest Pockels coefficient in BTO[5] is more than an order of magnitude larger than that for LN[6], yet the two materials still have similar optical properties which are determined from electronic resonances. Furthermore, the EO strength can be dispersive since the acoustic lattice vibrations dissipate between MHz and GHz frequencies, which is the range of interest to many of the applications mentioned in the introduction of the main text.

## 1.1 Factors Influencing the Frequency Dependence of the Permittivity in BTO

We attribute defects, domain structure and domain poling to be the main reasons behind the variation among the literature data in terms of the strength and the central component of the low frequency relaxation. The role of defects has been covered in the main text. Here, we examine the differences between single and poly domain crystals. Then we look the domain size and the effects of antiparallel and 90-degree ferroelectric polarizations (i.e. poling).

The difference between single and poly domain BTO is straightforward. In a single domain crystal, the pure *a*- or *c*-axis permittivity can be measured whereas in poly domain BTO the permittivity will always be a mix of *a*- and *c*-axis domains. Indeed, a previous study prepared BTO samples with varying fractions of 90-degree domains and measured the permittivity along the supposed *c*-axis[7]. They found that the permittivity increased in samples with more 90-degree domains than in those with only 180-degree domains because the 90-degree domains would contribute $\varepsilon_a$ rather than $\varepsilon_c$ to the effective permittivity. The opposite must then be true for measurements along the *a*-axis, where the presence of 90-degree domains reduces the permittivity due to an increased influence from $\varepsilon_c$. This might explain why Ref. 31 in Fig. 2d – the only one with a single domain sample that can measure pure $\varepsilon_a$ – starts with a rather high permittivity at low frequencies and then has a large drop.

As an example of the differences between poly domain crystals, consider the following. An *a*-axis BTO thin film like ours can have multiple domains where the *c*-axis is randomly oriented in one of four 90-degree orientations in the plane of the film. Along one of these 90-degree directions, half of the domains will be *c*-axis and the other half will be *a*-axis on average. In Supplementary Note 3, we show that in this case the effective permittivity should be an equal-weighted average of $\varepsilon_a$ and $\varepsilon_c$. Compare this to Refs. 33 and 35 of the main text which report data on a thin film and a ceramic, respectively. Both samples feature a mixture of domains where the *c*-axis is randomly oriented in all directions (i.e. not just in-plane). Now the effective permittivity is skewed towards $\varepsilon_a$ because the unit cells have two *a*-axes but only one *c*-axis. Since $\varepsilon_a \gg \varepsilon_c$, the permittivity will be larger in mixed *c*- and *a*-axis samples compared to an *a*-axis thin film like ours.

The variation in the relaxation frequency in the MHz-GHz is correlated with the domain size. Higher relaxation frequencies are associated with smaller domains (3 GHz for 230 nm vs. 770 MHz for 980 nm)[8]. This may also explain why thin films tend to have relaxations in the GHz range while 100s of MHz is more common for bulk crystals with larger dimensions and larger domains.

Finally, we discuss the poling of thin films. Neighbouring domains can have ferroelectric polarizations that are parallel, antiparallel or at right angles to one another. Clemens et al. also found that the permittivity drop along the c-axis tends to be greater when there are more 90-degree domains[7]. By the same logic as before, the drop along the a-axis should be smaller with more 90-degree domains. Antiparallel domains seem to have an even more significant influence[9]. Nakao et al. prepared BTO crystals with only 180-degree domains and then poled the crystals in stages to gradually remove the antiparallel domains. They found that the as-prepared crystals with random domain polarizations had a factor 3 higher $\varepsilon_c$ and a factor 2 higher $\varepsilon_a$ when compared to fully poled films that had no antiparallel domains. In addition, the raw films with many antiparallel domains showed almost no frequency dependence while the fully poled films showed a large permittivity drop around 1 MHz.

## 2 Extracting Permittivity from $S_{11}$ Reflection Measurements

The permittivity is found by fitting the impedance of an equivalent circuit model to the impedance measured with $S_{11}$ reflection measurements. The methods section of the main text describes the fitting process.

Presented here are the measured impedance and $S_{11}$ data as well as the equivalent circuit model that is used to analytically derive impedance and $S_{11}$. Using the circuit model one can determine the permittivity of the EO layer $\varepsilon_{\text{EO}}$ from a measurement of $S_{11}$. This is because the only free parameter of the impedance model is $\varepsilon_{\text{EO}}$. All other parameters that go into the model are fixed because they are related to either the geometry of the structure or the permittivity and conductivity of other materials in the structure.

We use $S_{11}$ reflection measurements because our phase shifters are small enough to effectively be a lumped element. Note that similar methods can be used for transmission lines if the phase shifters are longer[10,11].

## 2.1 Impedance Measurements and Models

To validate the impedance model, we use LN as a reference material. Literature data for LN's permittivity from many sources are in good agreement on the values of $\varepsilon_c$ = 27 and $\varepsilon_a$ = 45 (see Fig. 2(c) in the main text). Additionally, the literature agrees that these values should be constant over the measured frequency range. LN is a relatively simple case with few unknowns which makes it a good test case for the model.

The measured data for the real and imaginary parts of both $Z$ and $S_{11}$ are shown in Fig. S1 by dark blue circles. The gap in the data between 170 GHz and 220 GHz is a range that cannot be covered by our equipment. We then compare the measurements against $Z$ and $S_{11}$ values that are calculated from the equivalent circuit model (light blue lines). Towards this end we have used the values of $\varepsilon_c$ and $\varepsilon_a$ from literature and all other device geometry parameters were taken from the designed dimensions of the structure or commonly used material properties ($\varepsilon_{Si}$ = 11.7, $\varepsilon_{oxide}$ = 3.9, $\varepsilon_{PMMA}$ = 3.9, $\sigma_{Au}$ = 4×10$^8$ S/m). The circuit model, its parameters and details on how to calculate its elements are described in the next section. None of the model parameters were fitted to match the measure data. The excellent match between the model and the measured data suggests that the impedance model is well-founded. The only part of the measurements that is not a near-exact match is the real part of $Z$ at frequencies below 1 GHz. We presume it is the result of some dielectric loss that originates from dielectric resonances in the 1-100 MHz range[12,13].

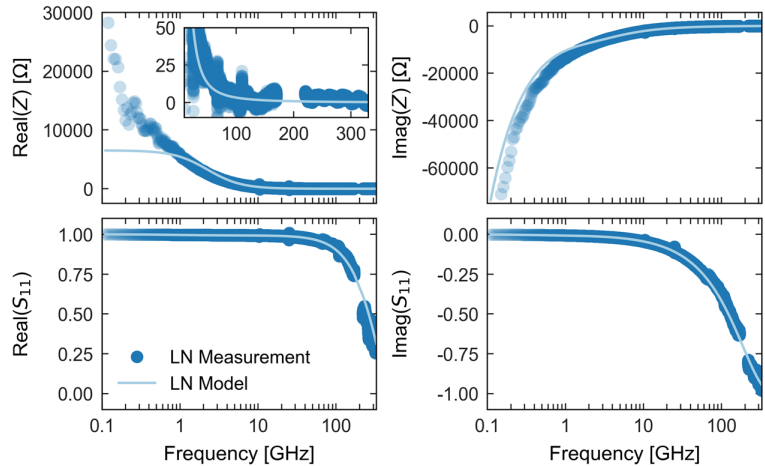

**Fig. S1 | Measured and modelled impedance data for LN.** The top row shows the real and imaginary parts of the impedance $Z$. The bottom row shows the real and imaginary parts of the $S_{11}$ reflection parameter.

The excellent match between the LN impedance measurements and the impedance calculated from the model using known parameters, prove that the impedance model is suitable for the structures in this work.

To add further confluence towards the model's accuracy, we use the same process for LN devices with three different lengths. The measured $S_{11}$ data and the $S_{11}$ predicted by the model are plotted in Fig. S2. For the longer devices at the highest frequencies, the lumped element assumption of the model starts to break down because the RF wavelength starts to approach the device length. This is why the phase shifters used for EO characterization are so short.

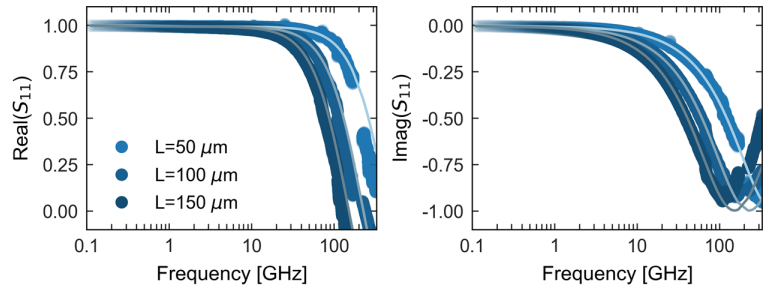

**Fig. S2 | Measured and modelled reflection data for LN devices with different lengths.** The real and imaginary parts of the $S_{11}$ reflection parameter are shown for LN devices with lengths of 50 μm, 100 μm and 150 μm. The EO characterization is performed with the 50 μm devices to stay within the lumped element approximation at higher frequencies.

Now we focus on the application of the model to BTO devices. Modelling the BTO devices is more challenging because $\varepsilon_{BTO}$ is expected to vary across the measured frequency range. Additionally, the dispersion of the real part of $\varepsilon_{BTO}$ also must come with an imaginary component. There is also no widespread agreement about $\varepsilon_{BTO}$ among previous reports as was the case for LN. Therefore, $\varepsilon_{BTO}$ must be found by fitting an analytical model to the data. We use the same impedance model for BTO that was used for LN but with geometric parameters adapted for the BTO devices. Instead of the constant permittivity model assumed for LN, we use the Debye model in equation (2) of the main text to get a frequency dependent $\varepsilon_{BTO}$. The Debye model takes four parameters as input: the central relaxation frequency $\gamma_0$, the standard deviation of relaxation frequencies in log-space $\sigma$ and the relaxation strength $S_R$. The high frequency permittivity is taken to be $\varepsilon_\infty$ = 231 based on the Debye model fitted to literature data (Table 1, main text). These three unknown variables are now used as fitting parameters in the impedance model. Fig. **S3** shows the measured BTO data in the dark green circles. The results calculated from the circuit model best fitting values for the Debye model parameters are plotted in the light green lines. The parameters that give the best fit to the data along with the standard errors of the fit are given in Table 1 of the main text. The BTO impedance model describes the measurements remarkably well and the small standard errors of the fit give us a high degree of confidence that BTO's permittivity in this frequency range is well-described by the Debye model.

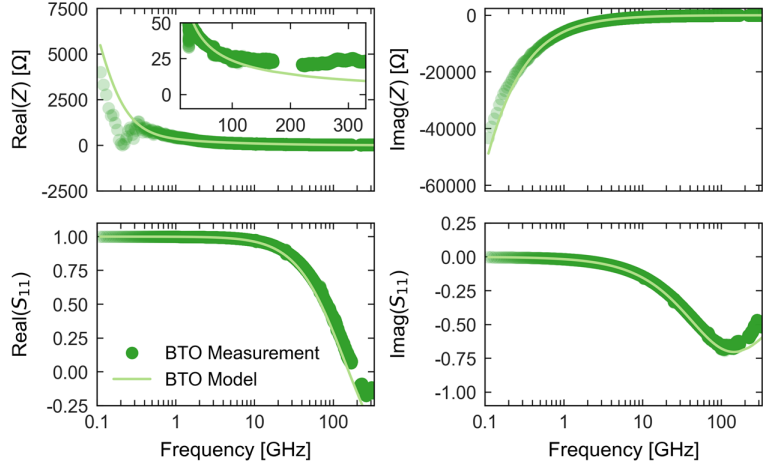

**Fig. S3 | Measured and modelled impedance data for BTO.** The top row shows the real and imaginary parts of the impedance $Z$. The bottom row shows the real and imaginary parts of the $S_{11}$ reflection parameter.

## 2.2 Equivalent Circuit Model of the Phase Shifter

This section describes the equivalent circuit model that was used to model the impedance of the phase shifters and subsequently extract the permittivity of the thin films. The calculation of the elements in the equivalent circuit will be described in the next section.

Fig. S4(a) shows a cross-sectional schematic of the phase shifter's layer stack overlaid with the various elements that contribute to the overall impedance. Between the signal and ground electrodes there exists a capacitance $C_{\mathrm{GSG}}$ which is the sum of partial capacitances due to each layer in the stack. In parallel to $C_{\mathrm{GSG}}$ is the conductance in the electro-optic (EO) layer $G_{\mathrm{EO}}$. The conductance represents the dielectric loss in the EO layer and for BTO it is essential for an accurate impedance model. Also in parallel to $C_{\mathrm{GSG}}$ and $G_{\mathrm{EO}}$ is the substrate impedance. The silicon substrate requires multiple elements to account for the conduction and displacement currents that depend on the modulation frequency[14,15]. At low frequencies substrate currents are dominated by carrier conduction and this is accounted for with $G_{\mathrm{si}}$. Displacement currents dominate at high frequencies and this is accounted for with the capacitive element $C_{\mathrm{si}}$. A second capacitor $C_{\mathrm{sub}}$ accounts for the capacitance between the signal electrode and the silicon substrate. Fig. S4(b) shows the arrangement of these elements in a simplified equivalent circuit model that highlights the parallel current paths. We note that here we show the full GSG electrode structure, in contrast to the figures of the main text where typically only one ground electrode is shown. The reason for this is that our high-frequency electrical probes have a GSG configuration. Fig. S4(c) shows how the phase shifters were incorporated into coplanar waveguide electrodes to be compatible with GSG probes. Each set of electrodes supports two phase shifters with one in each G-S gap, similar to Mach-Zehnder modulators. Unlike Mach-Zehnder modulators, however, each arm stays separate and has its own input and output grating couplers. This configuration has the advantage of doubling the number of phase shifters available to measure.

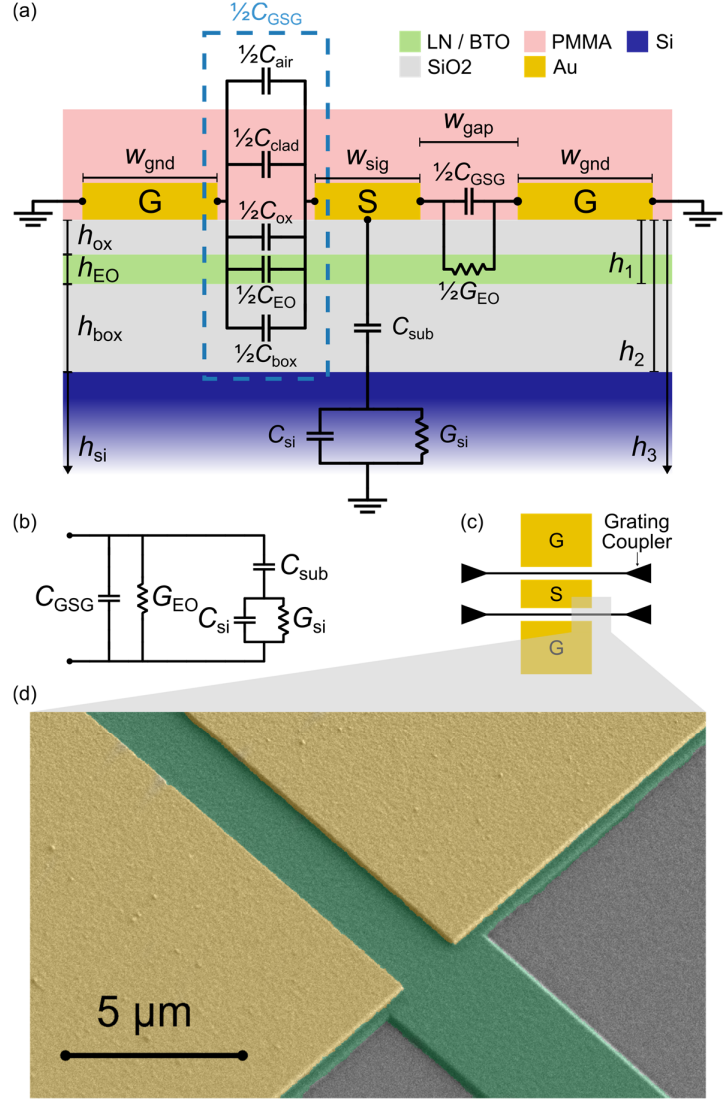

**Fig. S4 | Equivalent circuit model of the phase shifter. a**, Schematic of the various contributions to the total impedance of the phase shifter. **b**, Simplified Equivalent circuit model of the phase shifter. **c**, Illustration of the on-chip configuration of the phase shifters. Each GSG electrode supports two phase shifters, one in each G-S gap. This configuration is necessary to maintain compatibility with high-frequency electrical probes. **d**, Scanning electron microscope image of the transition from the waveguide to the start of the phase shifter.

### 2.3 Calculation of the Circuit Elements

This section describes the calculation of the individual circuit elements in Fig. S4.

### 2.3.1 $C_{\text{GSG}}$ and $C_{\text{si}}$ – Coplanar Waveguide Capacitance

The electrodes form a coplanar waveguide (CPW) which has a capacitance that can be calculated by conformal mapping methods[14,15]. For multi-layer substrates the capacitance of a CPW can be calculated with the following elliptic integrals[16]. The general form for the capacitance per unit length of a CPW with an effective permittivity $\varepsilon_{\text{eff}}$ is given by

$$C_{\text{CPW}} = 4\varepsilon_0 \varepsilon_{\text{eff}} \frac{\text{K}(k'_0)}{\text{K}(k_0)}, \tag{1}$$

where K is the complete elliptic integral of the first kind with modulus $k_0$ given by equation (2) and $k_0' = \sqrt{1 - k_0^2}$

$$k_0 = \frac{x_c}{x_b} \sqrt{\frac{x_c^2 - x_a^2}{x_b^2 - x_a^2}}. \tag{2}$$

The parameters $x_a$, $x_b$, $x_c$ are related to the CPW geometry

$$\begin{aligned} x_a &= \frac{w_{sig}}{2}, \\ x_b &= x_a + w_{gap}, \\ x_c &= x_b + w_{gnd}. \end{aligned} \tag{3}$$

If the CPW is embedded in a homogeneous medium then $\varepsilon_{eff} = \varepsilon_r$ where $\varepsilon_r$ is the permittivity of the medium. For multi-layer structures the partial capacitance technique allows one to calculate the capacitance contribution for each material layer individually. The total capacitance is then be given by the sum of the parallel partial capacitances. The calculation of the partial capacitances is modified from equation (1). Taking the silicon substrate in Fig. S4(a) as an example, one would first calculate the partial capacitance for a silicon layer that spanned the entire thickness $h_3$ and then subtract from that the partial capacitance of a silicon layer that spanned the thickness $h_2$. The remaining capacitance is that of the actual silicon layer. The elliptic integrals take a different form[16]

$$C_{si} = 2\varepsilon_0 \varepsilon_{si} \frac{K(k_3')}{K(k_3)} - 2\varepsilon_0 \varepsilon_{si} \frac{K(k_2')}{K(k_2)}, \tag{4}$$

with the moduli $k_i$ now given by equation (5), $k_i' = \sqrt{1 - k_i^2}$ and i = 1, 2, 3 corresponding to the thicknesses $h_i$ in Fig. S4(a)

$$k_i = \frac{\sinh\left(\frac{\pi x_c}{2h_i}\right)}{\sinh\left(\frac{\pi x_b}{2h_i}\right)} \sqrt{\frac{\sinh^2\left(\frac{\pi x_c}{2h_i}\right) - \sinh^2\left(\frac{\pi x_a}{2h_i}\right)}{\sinh^2\left(\frac{\pi x_b}{2h_i}\right) - \sinh^2\left(\frac{\pi x_a}{2h_i}\right)}}. \tag{5}$$

The factor 2 in equation (4) compared to the factor 4 in equation (1) occurs because the calculation for the silicon (or any other layer) is only for the half-space below the CPW.

Equations (4) and (5) can become inaccurate when the permittivity for each layer is not strictly decreasing away from the electrodes[17,18]. This is especially true for the layer stacks in this work where the permittivities of LN and BTO are much larger than that of the silicon dioxide layer that separates the EO materials from the CPW. Modifications to the conformal mapping formulas above have been proposed[17,18], however, we found it easier and more accurate to use EM simulations to calculate the effective permittivity $\varepsilon_{eff}$ of the CPW with the entire layer stack. This $\varepsilon_{eff}$ can then be used with equation (1) to calculate the total CPW capacitance. The dependence of $\varepsilon_{eff}$ on the permittivity of the EO layer $\varepsilon_{EO}$ is shown in Fig. S5(a).

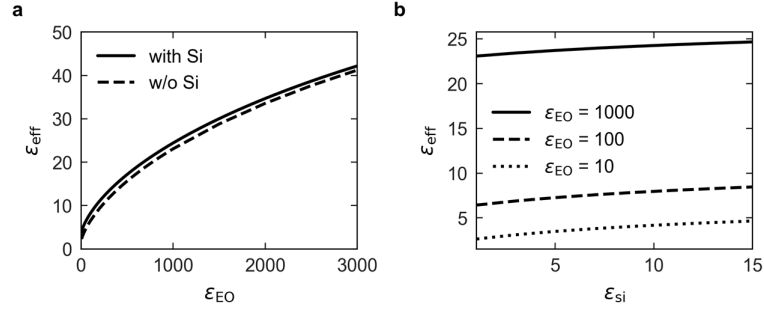

**Fig. S5 | Effective permittivity of the phase shifter. a**, CPW $\varepsilon_{\text{eff}}$ dependence structure on the EO permittivity $\varepsilon_{\text{EO}}$ with (solid) and without (dashed) the silicon substrate. **b**, CPW $\varepsilon_{\text{eff}}$ dependence on $\varepsilon_{\text{si}}$ for various values of $\varepsilon_{\text{EO}}$ to verify that the presence of the silicon layer does not strongly influence the total field distribution and that the method to determine $C_{\text{si}}$ is accurate.

Since $C_{\text{si}}$ is in parallel with $G_{\text{si}}$ it is still necessary to separate the capacitance due to the silicon layer from the capacitance due to the rest of the layer stack, which is labelled as $C_{\text{GSG}}$. To this end, we use a method similar to the idea of partial capacitances, where the capacitance of the full layer stack $C_{\text{CPW}}$ should be the sum of parallel capacitances such that $C_{\text{CPW}} = C_{\text{GSG}} + C_{\text{si}}$. First, the total capacitance $C_{\text{CPW}}$ is calculated based on the $\varepsilon_{\text{eff, CPW}}$ that is simulated with the full layer stack. Then $\varepsilon_{\text{eff, GSG}}$ is simulated without the silicon layer to calculate $C_{\text{GSG}}$. Finally, $C_{\text{si}}$ is given by the difference $C_{\text{CPW}} - C_{\text{GSG}}$. The effective permittivity with and without the silicon layer is shown in Fig. S5(a) by the solid and dashed lines, respectively. To verify that removing the silicon layer doesn't strongly influence the field distribution and thus, the value $\varepsilon_{\text{eff, GSG}}$, $C_{\text{si}}$ was calculated using various values for $\varepsilon_{\text{si}}$ in the simulation. The capacitance should be linearly proportional to $\varepsilon_{\text{si}}$ if the rest of the rest of the geometry is fixed. Fig. S5(b) shows that this is approximately true. Note that all capacitances in this section are per unit length, so the actual capacitance of the device is obtained by multiplying by the length of the CPW.

### 2.3.2 $G_{\text{EO}}$ – Dielectric Loss in the EO Layer

Dielectric loss occurs whenever a material's permittivity has an imaginary component. In general, when the real part of the permittivity changes there will also be an imaginary component as per the Kramers-Kronig relations. For LN this is not an issue in the measured frequency range. For BTO, the Debye model in equation (2) of the main text provides the complex permittivity $\varepsilon' + j\varepsilon''$ and the loss tangent can be calculated as $\tan \delta = \varepsilon''/\varepsilon'$. The dielectric loss in the BTO layer can be modelled as a conductance $G_{\text{EO}}$ in parallel to $C_{\text{GSG}}$. Since the dielectric loss in the other layers is negligible compared to the loss in the BTO layer, the entire loss (or conductivity) must come from the active electro-optical structure, i.e. $G_{\text{EO}} \approx G_{\text{GSG}}$. Dielectric loss can be calculated with the same elliptic integrals as in the previous sections[19]. In fact, the dielectric loss is simply the capacitance multiplied by a factor $2\pi f \cdot \tan \delta$ which means that $G_{\text{EO}}$ in the context of this analysis is given by

$$G_{\text{EO}} = 2\pi f \cdot \tan \delta \cdot C_{\text{GSG}} \ . \tag{6}$$

$G_{\text{EO}}$ is essential to the impedance model if it is to match the real and imaginary parts of the measured impedance simultaneously. Anecdotally, if $G_{\text{EO}}$ is excluded from the impedance model then $\varepsilon_{\text{BTO}}$ will typically be overestimated, which would also lead to overestimated values for the Pockels coefficient. Note that equation (6) gives $G_{\text{EO}}$ per unit length and must therefore be multiplied by the length of the CPW to get the dielectric loss of an actual device.

### 2.3.3 $G_{si}$ – Silicon Conductance

The silicon conductance is calculated with conformal mapping in a similar way to its partial capacitance[15]

$$G_{si} = \frac{2}{\rho_{si}} \frac{K(k_3')}{K(k_3)} - \frac{2}{\rho_{si}} \frac{K(k_2')}{K(k_2)}, \tag{7}$$

where $\rho_{si}$ is the silicon resistivity and $k_i$, $k_i'$ are given by equation (5). Once again, the conformal mapping method is not the most accurate for the layer stack in this work. However, $G_{si}$ is so small for the high-resistivity silicon substrates in this work that $G_{si}$ – and the inaccuracy of its calculation – is irrelevant to the impedance over the measured frequency range. If an accurate calculation of $G_{si}$ is required, we propose the following method to calculate a correction term for the elliptic integrals. First, calculate the approximated silicon layer capacitance $C_{si, approx.}$ according to equation (4). Then, calculate the actual silicon layer capacitance $C_{si}$ using electromagnetic simulations as described in section 2.2.1. Finally, the ratio $\kappa = C_{si}/C_{si, approx.}$ gives a scaling factor that corresponds to how much the elliptic integrals under-/overestimate the actual quantity. Keeping the same notation where $G_{si, approx.}$ is given by equation (7), then the actual silicon conductance would be given by $G_{si} = \kappa G_{si, approx.}$. Note that equation (7) gives $G_{si}$ per unit length and must therefore be multiplied by the length of the CPW to get the silicon conductance of an actual device.

### 2.3.4 $C_{sub}$ – Signal-Substrate Capacitance

The signal-substrate capacitance is given by the standard parallel plate capacitance model

$$C_{pp} = \varepsilon_0 \varepsilon_r \frac{w_{sig} l}{h}, \tag{8}$$

where $l$ is the length of the CPW. The parallel plate capacitance is calculated for each of the layers between the signal electrode and the silicon substrate. The total signal-substrate capacitance is then

$$\begin{aligned} C_{sub} &= \left( C_{sub, ox}^{-1} + C_{sub, EO}^{-1} + C_{sub, box}^{-1} \right)^{-1} \\ &= \left( \left( \varepsilon_0 \varepsilon_{ox} \frac{w_{sig} L}{h_{ox}} \right)^{-1} + \left( \varepsilon_0 \varepsilon_{EO} \frac{w_{sig} L}{h_{EO}} \right)^{-1} + \left( \varepsilon_0 \varepsilon_{box} \frac{w_{sig} L}{h_{box}} \right)^{-1} \right)^{-1}. \end{aligned} \tag{9}$$

Since these capacitances are in series, $C_{sub}$ is dominated by the smallest capacitor which is the buried oxide layer. Since $h_{box} \gg h_{EO}$ and $\varepsilon_{box} \ll \varepsilon_{EO}$, there is a negligible dependence on $C_{sub, EO}$ and therefore $\varepsilon_{EO}$ as well.

### 2.3.5 Longitudinal Impedance

Most analyses of CPW impedance also include elements to account for longitudinal currents in the structure[15,19,20]. This includes the inductance and resistance of the electrodes, as well as the resistance of the silicon substrate in the propagation direction. These elements would be placed in series with the simplified equivalent circuit of Fig. **S4**(b). For the CPWs in this work, however, the device lengths are so small that the longitudinal elements can be neglected for the measured frequency range. For completeness, details on the calculation of these elements are provided below.

The inductance of the electrodes $L_{\text{CPW}}$ is given by conformal mapping with $k_0$ given by[14,20]

$$L_{\text{CPW}} = \frac{4}{\mu_0} \frac{K(k_0)}{K(k_0')}, \qquad (10)$$

where $\mu_0$ is the permeability of free space. The resistance along the signal electrode $R_{\text{sig}}$ depends on the skin depth $\delta$ of the electrode. The electrodes in our devices are only 200 nm thick and are much smaller than the skin depth except for frequencies above 300 GHz. Regardless, the skin depth can be calculated by[21]

$$\delta = \sqrt{\frac{2}{\sigma \omega \mu}} \sqrt{\sqrt{1 + \left(\frac{\omega \varepsilon}{\sigma}\right)^2} + \frac{\omega \varepsilon}{\sigma}}, \qquad (11)$$

where $\omega$ is the angular frequency, $\sigma$ is the material's conductivity, $\mu = \mu_0 \mu_r$ and $\varepsilon = \varepsilon_0 \varepsilon_r$. When the thickness of the electrodes $t \leq \delta$ then

$$R_{\text{sig}} = \frac{1}{\sigma_{\text{sig}} \cdot t \cdot w_{\text{sig}}}, \qquad (12)$$

where $\sigma_{\text{sig}}$ is the conductivity of the electrode. When $t \geq \delta$ then $R_{\text{sig}}$ depends on the skin depth – and therefore frequency as well – and is given by

$$R_{\text{sig}} = \frac{1}{\sigma_{\text{sig}} \cdot \delta_{\text{sig}} \cdot w_{\text{sig}}}. \qquad (13)$$

Similarly, the longitudinal resistance in the silicon substrate is given by

$$R_{\text{L}} = \frac{\rho_{\text{si}}}{\delta_{\text{si}} \cdot w_{\text{sig}}}. \qquad (14)$$

Where $\rho_{\text{si}}$ is the resistivity of the silicon. $L_{\text{CPW}}$ and $R_{\text{sig}}$ are connected in series while $R_{\text{L}}$ offers a parallel current path to the other two. Typically $R_{\text{L}}$ is much greater than the impedance of $R_{\text{sig}}$ and $L_{\text{CPW}}$ and is therefore inconsequential.

# 3 Extracting Pockels Coefficients From Phase Modulation Measurements

## 3.1 Derivation of the Angular Dependence of $r_\text{eff}$

Here we derive the analytical formula for the effective Pockels coefficient[22,23]. We start with the Pockels tensor for LN because it contains the most elements, including all elements present in BTO. Fig. S6 illustrates the coordinate system (gray arrows) and the various angles that will be used in the following derivation.

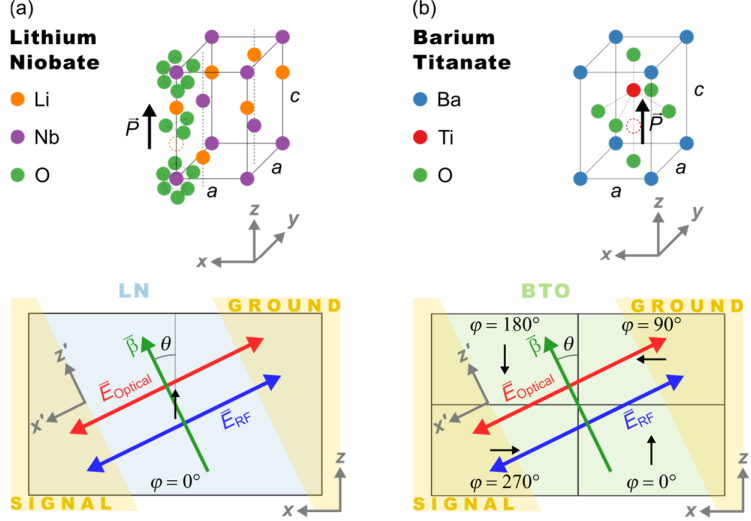

**Fig. S6: Coordinate system for LN (a) and BTO (b) relative to the crystal axes.** The original coordinate system is given by $x, y, z$, which are marked with the gray arrows. The rotated coordinate system corresponding to principal axes of the modified index ellipsoid under the influence of an external RF field is described with $x', y', z'$.

The refractive index in a crystal aligned where the $x, y, z$ directions are aligned with the $a, b, c$ crystal axes as in Fig. S6 can be described by an index ellipsoid[24]

$$1 = \left[\frac{1}{n_1^2} + \Delta\left(\frac{1}{n^2}\right)_1\right]x^2 + \left[\frac{1}{n_2^2} + \Delta\left(\frac{1}{n^2}\right)_2\right]y^2 + \left[\frac{1}{n_3^2} + \Delta\left(\frac{1}{n^2}\right)_3\right]z^2 \\ + 2\Delta\left(\frac{1}{n^2}\right)_4 yz + 2\Delta\left(\frac{1}{n^2}\right)_5 xz + 2\Delta\left(\frac{1}{n^2}\right)_6 xy. \quad (15)$$

The change in refractive index along a given dimension is given by[24]

$$\Delta\left(\frac{1}{n^2}\right)_{ij} = \sum_{k=1}^{3} r_{ijk} E_{\text{RF},k}, \quad (16)$$

where $E_{\text{RF},k}$ is the RF electric field along direction $k = x, y, z$. The matrix form including the electro-optic coefficients of LN is given by

$$\begin{bmatrix} \Delta(1/n^2)_1 \\ \Delta(1/n^2)_2 \\ \Delta(1/n^2)_3 \\ \Delta(1/n^2)_4 \\ \Delta(1/n^2)_5 \\ \Delta(1/n^2)_6 \end{bmatrix} = \begin{bmatrix} 0 & -r_{22} & r_{13} \\ 0 & r_{22} & r_{13} \\ 0 & 0 & r_{33} \\ 0 & r_{42} & 0 \\ r_{42} & 0 & 0 \\ -r_{22} & 0 & 0 \end{bmatrix} \begin{bmatrix} E_{\text{RF},x} \\ E_{\text{RF},y} \\ E_{\text{RF},z} \end{bmatrix}, \quad (17)$$

where we have used the reduced Voigt notation for indices ij. The situation for BTO is obtained by setting $r_{22}$ = 0. Carrying out the matrix multiplication leads to

$$\begin{bmatrix} \Delta(1/n^2)_1 \\ \Delta(1/n^2)_2 \\ \Delta(1/n^2)_3 \\ \Delta(1/n^2)_4 \\ \Delta(1/n^2)_5 \\ \Delta(1/n^2)_6 \end{bmatrix} = \begin{bmatrix} r_{13}E_{\text{RF},z} - r_{22}E_{\text{RF},y} \\ r_{13}E_{\text{RF},z} + r_{22}E_{\text{RF},y} \\ r_{33}E_{\text{RF},z} \\ r_{42}E_{\text{RF},y} \\ r_{42}E_{\text{RF},x} \\ -r_{22}E_{\text{RF},y} \end{bmatrix}. \tag{18}$$

Now that we have the $\Delta(1/n^2)$ for each set of ij indices, the index ellipsoid can be written in the form

$$\begin{aligned} 1 = &\left(\frac{1}{n_o^2} - r_{22}E_{\text{RF},y} + r_{13}E_{\text{RF},z}\right)x^2 + \left(\frac{1}{n_o^2} + r_{22}E_{\text{RF},y} + r_{13}E_{\text{RF},z}\right)y^2 \\ &+ \left(\frac{1}{n_e^2} + r_{33}E_{\text{RF},z}\right)z^2 \\ &+ (r_{42}E_{\text{RF},y})2yz + (r_{42}E_{\text{RF},x})2xz - (r_{22}E_{\text{RF},y})2xy\,. \end{aligned} \tag{19}$$

In our case, the ordinary axis is the *a*-axis with $n_1 = n_2 = n_o$ and the extraordinary axis is the *c*-axis with $n_3 = n_e$. For the *a*-axis substrate the *z*- and *x*-axes lie in-plane and the *y*-axis is out of the plane. The waveguide is oriented along the *z*-axis and both the optical and external fields are polarized along the *x*-axis when $\theta = 0$. To find the optimum angle $\theta$ we first apply a coordinate rotation around the *y*-axis to get new axes $z'$ and $x'$

$$\begin{pmatrix} x \\ z \end{pmatrix} = \begin{pmatrix} \cos\theta & \sin\theta \\ -\sin\theta & \cos\theta \end{pmatrix} \begin{pmatrix} x' \\ z' \end{pmatrix}. \tag{20}$$

The axes and the electric fields can be expressed as

$$\begin{aligned} x &= x'\cos\theta + z'\sin\theta \\ y &= y' \\ z &= z'\cos\theta - x'\sin\theta \\ E_{\text{RF},x} &= E_{\text{RF},x'}\cos\theta \\ E_{\text{RF},z} &= -E_{\text{RF},z'}\sin\theta\,. \end{aligned} \tag{21}$$

Here we have omitted electric field terms in the $y'$ and $z'$ directions because both optical and RF fields in the EO region of the phase shifter are only in the $x'$ direction. Each of these terms is inserted into the index ellipsoid. After expanding all multiplication terms, we can drop all terms except those that contain $x'x' = x'^2$ because we are confined to the case where the optical and RF fields are both polarized along the $x'$ direction. The resulting index ellipsoid is reduced to

$$1 = \left[\frac{\cos^2\theta}{n_o^2} + \frac{\sin^2\theta}{n_e^2} - r_{13}E_{\text{RF},x'}\cos^2\theta\sin\theta - r_{33}E_{\text{RF},x'}\sin^3\theta \\ - 2r_{42}E_{\text{RF},x'}\cos^2\theta\sin\theta\right]x'^2\,. \tag{22}$$

The refractive index along the $x'x'$ direction is then given by

$$\frac{1}{n_{x'x'}^2} = \left[\frac{\cos^2\theta}{n_o^2} + \frac{\sin^2\theta}{n_e^2} - (r_{13} + 2r_{42})E_{\text{RF},x'}\cos^2\theta\sin\theta \\ - r_{33}E_{\text{RF},x'}\sin^3\theta\right]. \tag{23}$$

In the absence of an external field, the refractive index of the $x'x'$-polarized wave is a combination of $n_o$ and $n_e$

$$\frac{1}{n_{x'x'}^2}(\theta) = \frac{\cos^2\theta}{n_o^2} + \frac{\sin^2\theta}{n_e^2} \quad \text{or} \quad n_{x'x'}(\theta) = \frac{n_o n_e}{\sqrt{n_o^2 \sin^2\theta + n_e^2 \cos^2\theta}} \ . \tag{24}$$

In the presence of an external field $E_{\Omega,x'}$ the remaining terms give the change in refractive index

$$\Delta\left(\frac{1}{n_{x'x'}^2}\right) = -E_{RF,x'}[(r_{13} + 2r_{42})\cos^2\theta \sin\theta + r_{33}\sin^3\theta] \ . \tag{25}$$

The term in brackets is what determines the effective Pockels coefficient $r_{\text{eff}}$ which in this case is equivalent to $r_{x'x'}$.

$$r_{\text{eff}}(\theta) = (r_{13} + 2r_{42})\cos^2\theta \sin\theta + r_{33}\sin^3\theta \tag{26}$$

For BTO, one must also consider the effect of multiple domains oriented at 90° to one another. For this, we introduce $\varphi$ to describe the domain orientation ($\varphi$ = 0°, 90°, 180°, 270°) as well as $\nu_\varphi$ to describe the fraction of domains with orientation $\varphi$ ($0 \leq \nu_\varphi \leq 1$ and $\sum_\varphi \nu_\varphi = 1$). The effective Pockels coefficient of the multi-domain film is then the weighted average of $r_{\text{eff}}(\theta + \varphi)$ over all domain orientations $\varphi$ with weights given by $\nu_\varphi$.

$$r_{\text{eff}}(\theta) = \sum_\varphi \nu_\varphi (\cos^2(\theta + \varphi) \sin(\theta + \varphi)(r_{13} + 2r_{42}) + r_{33}\sin^3(\theta + \varphi)) \tag{27}$$

### 3.1.1 Angular Dependence of the Permittivity

The permittivity in the BTO film has no angular dependence because it is an average across many domains oriented at right angles to each other. In a single domain, the angular dependence of the refractive index can be described by the radius of the ellipse with semi-axes corresponding to the refractive index along each crystal axis

$$n^2(\theta) = n_c^2 \sin^2\theta + n_a^2 \cos^2\theta \tag{28}$$

The net refractive index of the film is the average across all domains. The refractive index is the same for anti-parallel domains so only 90-degree domains need to be considered. The net refractive index is then

$$\begin{aligned} n^2(\theta) &= \frac{1}{2}[(n_c^2 \sin^2\theta + n_a^2 \cos^2\theta) + (n_a^2 \sin^2\theta + n_c^2 \cos^2\theta)] \\ n^2(\theta) &= \frac{(n_a^2 + n_c^2)}{2}[\sin^2\theta + \cos^2\theta] = \frac{(n_a^2 + n_c^2)}{2} \end{aligned} \tag{29}$$

With $\varepsilon = n^2$, the permittivity is $\varepsilon = \frac{1}{2}(\varepsilon_a + \varepsilon_c)$. Thus, there is no angular dependence for multi-domain films.

### 3.2 Fitting $r_{ij}$ Coefficients to the Measured $r_{\text{eff}}$ Data

In this section we describe the fitting process that was used to extract the individual Pockels tensor elements from the measurements of the effective Pockels coefficient. While the effective Pockels coefficient is useful as a single number that quantifies the EO strength in a device, it is still of interest to know the values of the individual elements that comprise the Pockels tensor. The angular dependence of $r_{\text{eff}}$ provides a framework to do this. Equation (27) indicates a dependence on $\theta$ with coefficients $r_{13} + 2r_{42}$ and $r_{33}$. This assumes fixed values of $\varphi$ and $\nu_\varphi$ which is the case for our measurements since we initially apply a DC bias and then remove it before allowing the films to reach an equilibrium state for the domain fractions (see Supplementary Note S2.4). For LN we have $\nu_{\varphi=0°} = 1$ because it is a single domain film. For BTO we use the partial-poled cased described above with $\nu_{\varphi=0°,90°} = 0.375$ and $\nu_{\varphi=180°,270°} = 0.125$. These domain fractions best represent the situation where the domain fractions are allowed to reach an equilibrium state after an initial poling step, which is described in the following section[25].

If one has a collection of $r_{\text{eff}}(\theta)$ measurement points from enough angles, then this equation can be fit to the data by using the aforementioned coefficients as fitting parameters. In this way, individual Pockels tensor elements can be determined rather than just $r_{\text{eff}}$. We take $r_{\text{eff}}(\theta, p_1, p_2)$ where $\theta$ is the independent variable and the new parameters are $p_1 = r_{13} + 2r_{42}$ and $p_2 = r_{33}$. Fig. S7 shows the measured $r_{\text{eff}}(\theta)$ points at a frequency of 10 GHz as well as the fitted curve according to equation (27). The LN data in Fig. S7 was fitted with values of $p_1$ = 28.9 ± 2.0 pm/V and $p_2$ = 26.9 ± 0.9 pm/V while the BTO data was fitted values of $p_1$ = 838 ± 29 pm/V and $p_2$ = 64.8 ± 16.2 pm/V. For LN which is a single domain thin film, there is only one non-zero domain fraction $\nu_{\varphi=0°} = 1$. For BTO we used domain fractions of $\nu_{\varphi=0°,90°} = 0.375$ and $\nu_{\varphi=180°,270°} = 0.125$, corresponding to the equilibrium domain fractions as explained in the main text. The domain fractions for BTO lead to an equation (27) that has symmetry every 45°. For LN, the resulting form of equation (27) leads to symmetry for every 90°. The LN devices with $\theta < 20°$ had modulation signals that were too weak measure, because the effective Pockels coefficient at these angles is so small.

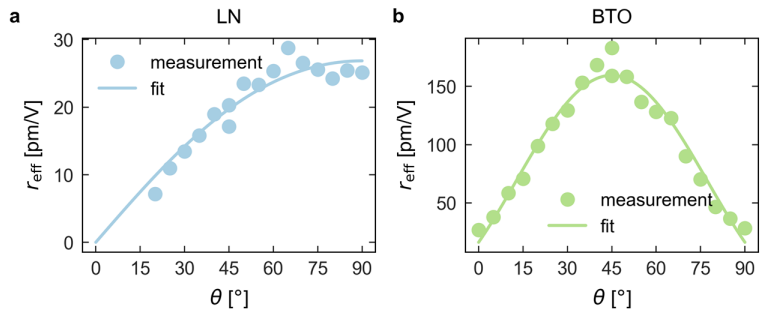

**Fig. S7 | Angular dependence of the effective Pockels coefficient.** Measured effective Pockels coefficients (circles) and the resulting fit (line) of equation (27) to the measurement points for **a** LN and **b** BTO. The plots show the results measured at a modulation frequency of 10 GHz. This process was repeated for each frequency point the get the results in Fig. 4 of the main text.

### 3.3 Poling Dynamics

Here we discuss the procedure for aligning the ferroelectric polarization of a majority of BTO domains in the same orientation to achieve a non-zero net EO effect (see the discussion of random vs. partial-poled vs. poled in the main text). Note that LN does not require poling because the thin film is already in a poled, single-domain state.

For each BTO device an initial poling procedure was performed where a DC bias was applied in addition to a 50 GHz modulation signal. The difference in dB between the powers in the first modulation sideband and in the optical carrier (peak-sideband ratio) was monitored while the magnitude of the DC bias was increased in 1 V increments. The sideband powers increase with the bias voltage as BTO domains start to flip polarizations. Eventually the peak-sideband ratio saturates at a maximum when no more domain polarizations are able to switch even with further increases in the bias. This corresponds to the "poled" state. Ideally, all electrical and EO measurements would be performed under this maximum bias, however, this was not possible for frequencies above 110 GHz due to available equipment. For consistency across all frequencies, all measurements were therefore performed without a DC bias.

To ensure a stable domain polarization state, the time decay of the peak-sideband ratio was measured. When the bias is removed, domains start to randomly flip to the opposite polarization and the peak-sideband ratio immediately decrease. Fig. S8 shows the decay in the peak-sideband ratio (i.e. modulation efficiency) over time and measured from the instant that the bias was removed. The modulation efficiency eventually settles to a value around 1 dB less than its initial value. This corresponds to the "partial-poled" state, where a majority of domains are still aligned in a preferred direction. All electrical and EO measurements were performed after this poling and decay sequence.

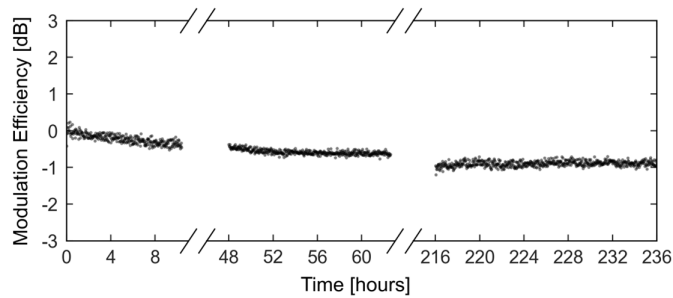

**Fig. S8 | Time decay of modulation efficiency.** Modulation efficiency as a function of time since the initial DC bias was removed that was used to pole the BTO film. The modulation efficiency stabilizes at a value only 1 dB less than the maximum efficiency with the DC bias.

## 4 Derivation of the Dependence of $\Delta n$ on $\varepsilon_{\text{BTO}}$

This section provides a derivation for equations (6-8) in the main text, which are analytical formulas for the refractive index change of an EO material in the context of the hypothetical structure in Fig. S9.

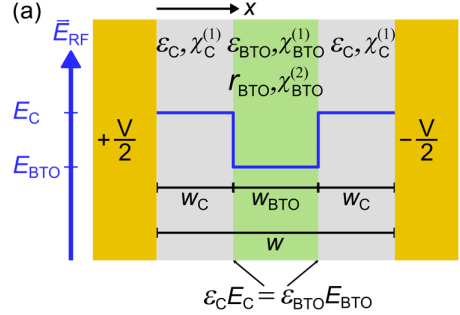

**Fig. S9 | Hypothetical phase shifter used to find an analytical dependence of the refractive index change in BTO layer on the susceptibility or permittivity of the materials in the structure.** A potential difference between metal electrodes denoted with G and S results in an electric field $\vec{E}_{\text{RF}}$ (blue arrow) across the layers. The middle layer in green is BTO. It is separated from the electrodes by a cladding material (gray).

The Pockels effect depends on the electric field $E_{\text{BTO}}$ in the region of the BTO. Therefore, we start by looking for an expression for $E_{\text{BTO}}$ that depends on the permittivities $\varepsilon_{\text{BTO}}$ and $\varepsilon_{\text{C}}$ as well as the layer widths $w_{\text{BTO}}$ and $w_{\text{C}}$. The potential difference $V$ between the ground and signal electrodes is

$$V = \int_0^{2w_C} E_C dx + \int_0^{w_{\text{BTO}}} E_{\text{BTO}} dx \tag{30}$$
$$= 2w_C E_C + w_{\text{BTO}} E_{\text{BTO}} \ .$$

Rearranging for $E_{\text{BTO}}$ gives

$$E_{\text{BTO}} = \frac{V - 2w_C E_C}{w_{\text{BTO}}} = \frac{V}{w_{\text{BTO}}} - \frac{2w_C E_C}{w_{\text{BTO}}} \ . \tag{31}$$

Let $R_w = 2w_C/w_{\text{BTO}}$ be the ratio of the total width of the cladding material to the width of the EO such that the total width $w$ between the electrodes can be expressed as $w = 2w_C + w_{\text{BTO}} = w_{\text{BTO}}(1 + R_w)$. This allows equation (31) to be rewritten as

$$E_{\text{BTO}} = \frac{V}{w}(1 + R_w) - R_w E_C$$
$$= \frac{V}{w}(1 + R_w) - R_w \frac{\varepsilon_{\text{BTO}}}{\varepsilon_C} E_{\text{BTO}} \ , \tag{32}$$

where $E_C$ was eliminated by using the boundary condition $\varepsilon_C E_C = \varepsilon_{\text{BTO}} E_{\text{BTO}}$ for electric fields that are perpendicular to an interface between two materials. Equation (32) can then be solved to find the expression for $E_{\text{BTO}}$ that was sought from the start

$$E_{\text{BTO}} = \frac{V}{w} \frac{1 + R_w}{1 + R_w \frac{\varepsilon_{\text{BTO}}}{\varepsilon_C}} \ . \tag{33}$$

In the second step we relate the first and second order susceptibilities. Miller's rule in equation (18) of the main text provides this relation

$$\delta = \frac{\chi^{(2)}(\omega_0 \pm \omega_m, \omega_0, \omega_m)}{\chi^{(1)}(\omega_0 \pm \omega_m)\chi^{(1)}(\omega_0)\chi^{(1)}(\omega_m)} \ . \tag{34}$$

The optical frequencies are much larger than the modulating frequencies so $\omega_0 \pm \omega_m \approx \omega_0$ and $\chi^{(1)}(\omega_0 \pm \omega_m) \approx \chi^{(1)}(\omega_0)$. The second order susceptibility as a function of the first order RF susceptibility is then

$$\chi^{(2)}(\omega_0 \pm \omega_m, \omega_0, \omega_m) = \delta \left(\chi^{(1)}(\omega_0)\right)^2 \chi^{(1)}(\omega_m) \ . \tag{35}$$

For the final step we find an expression for the change of refractive index in the electro-optic material $\Delta n$ that depends on $\varepsilon_{BTO}$ rather than on $r_{BTO}$. Start with the common expression for $\Delta n$ given by

$$\Delta n = -\frac{1}{2} n_0^3 r_{BTO} E_{BTO} \ . \tag{36}$$

where $n_0$ represents the value of the unperturbed optical refractive index. Using the fact that $r_{BTO} = -2\chi^{(2)}_{BTO}/n^4$, where $\chi^{(2)}_{BTO} = \chi^{(2)}(\omega_0 \pm \omega_m, \omega_0, \omega_m)$, we can write

$$\Delta n = \frac{\chi^{(2)}_{BTO}}{n_0} E_{BTO} \ . \tag{37}$$

Now inserting the expressions of equation (33) and equation (35) gives

$$\Delta n = \frac{1}{n_0}\left[\delta\left(\chi^{(1)}(\omega_0)\right)^2 \chi^{(1)}_{BTO}(\omega_m)\right]\left[\frac{V}{w}\frac{1+R_w}{1+R_w\frac{\varepsilon_{BTO}}{\varepsilon_C}}\right] \ . \tag{38}$$

We would like the expression to be normalized to a given $V/w$. Also, the optical refractive index $n_0$ and susceptibility $\chi^{(1)}(\omega_0)$ are taken to be constant. These terms are therefore grouped with $\delta$ into a single constant $\eta$ such that

$$\Delta n = \eta \frac{(1+R_w)\chi^{(1)}_{BTO}(\omega_m)}{1+R_w \frac{\varepsilon_{BTO}}{\varepsilon_C}},$$

$$\eta = \delta \frac{\left(\chi^{(1)}(\omega_0)\right)^2}{n_0}\frac{V}{w} = \delta\frac{(n_0^2-1)^2}{n_0}\frac{V}{w}. \tag{39}$$

Finally, we use the fact that $\chi^{(1)}_{BTO}(\omega_m) = \varepsilon_{BTO} - 1$ to arrive at the expression in equation (6) of the main text

$$\Delta n = \eta \frac{(1+R_w)(\varepsilon_{BTO}-1)}{1+R_w\frac{\varepsilon_{BTO}}{\varepsilon_C}} \ . \tag{40}$$

In the limit of small $R_w$ (i.e. plasmonic devices, $w_C = 0$), equation (40) can be simplified to equation (7) in the main text

$$\Delta n_{BTO}(R_w = 0) = \eta(\varepsilon_{BTO}-1) \ . \tag{41}$$

In the limit of large $R_w$ (i.e. photonic devices, $w_C \gg w_{BTO}$), equation (40) can be simplified to the approximation given in equation (8) in the main text

$$\Delta n_{BTO}(R_w \gg 1) \approx \eta\varepsilon_C \frac{\varepsilon_{BTO}-1}{\varepsilon_{BTO}}. \tag{42}$$

## 5 Voltage-Length Product Dependencies

In this section we discuss the dependence of the phase shifter's voltage-length product $V_\pi L$ on various structural parameters and the modulation frequency, as calculated from simulations.

### 5.1 Varying Ferroelectric Permittivity

Here we discuss the $V_\pi L$ dependence on the permittivity of the EO layer $\varepsilon_{EO}$ and the modulation frequency. The last section of the main text describes the relationship between the Pockels shift and the permittivity of the EO material. Specifically, it discusses how the RF electric field is pushed out of the EO layer when $\varepsilon_{EO}$ is large, which should reduce the modulator's efficiency. In Fig. S10(a) we show this effect for the phase shifter in this work. The $V_\pi L$ is calculated from simulations according to the procedure described in the methods section. The effective Pockels coefficient is held constant for all $\varepsilon_{EO}$ to highlight the effect of the changing RF electric field distribution. A $r_{eff}$ of 440 pm/V was chosen since it is the maximum value of $r_{eff}$ in the fully poled case (see Fig. 3, main text). Fig. S10(a) shows that the $V_\pi L$ increases (i.e. modulator becomes less efficient) for higher $\varepsilon_{EO}$ as a result of the smaller RF electric field in the EO region. This result also highlights the importance of having accurate permittivity measurements when it comes to measuring $V_\pi L$ and Pockels coefficients.

Simulations of the $V_\pi L$ also show that there is little dependence on the modulation frequency in the range of interest to this work. Fig. S10(b) shows this dependence from 100 MHz to 500 GHz for various values of $\varepsilon_{EO}$. This indicates that the cross-section of the phase shifter is small enough relative to the RF wavelengths such that the RF electric field distribution is not changing with frequency. In other words, the quasi-static TEM assumption that is assumed for the conformal mapping equations in section 2 above seems to hold up to at least 500 GHz.

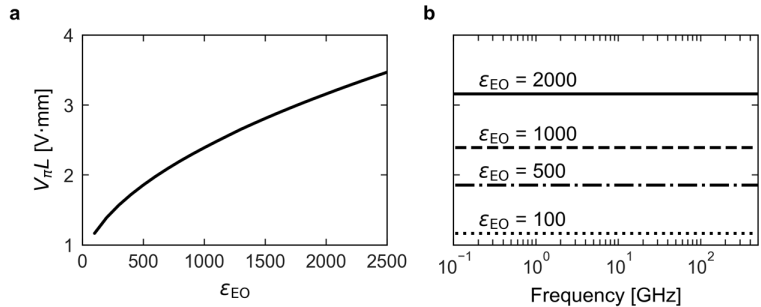

**Fig. S10 | Dependence of the voltage-length product on permittivity and frequency. a**, Simulated voltage-length product of the phase shifter in this work for various values of the EO layer's permittivity and assuming a constant $r_{eff}$ = 440 pm/V. **b**, Simulated voltage-length product of the phase shifter in this work as a function of modulation frequency and for various values of EO layer permittivity.

### 5.2 Varying Layer Thicknesses

The main text alluded to the phase shifters in this work having an optimal design for EO characterization. Part of the reason for this is the reproducibility of the fabrication process and the minimal uncertainty in the dimensions of the devices under test. One of the few areas of uncertainty that can influence the measured results is the thickness of the EO and oxide layers. Fig. S11(a) shows the simulated dependence of the phase shifter's voltage-length product on the error in height of the EO layer $\Delta h_{EO}$. The $V_\pi L$ becomes smaller for negative $\Delta h_{EO}$ (i.e. smaller $h_{EO}$) due to changes in the distributions of the optical and RF modes. Simulations were performed with different values for the permittivity of the EO layer $\varepsilon_{EO}$ to account for the permittivity dispersion of BTO, however, this has little influence. Two values of $\varepsilon_{EO}$ are plotted in Fig. S11 to show this, roughly corresponding to the values for LN ($\varepsilon_{EO}$ = 30, blue) and BTO at low frequency ($\varepsilon_{EO}$ = 2500, green). The data is plotted across a range covering $\Delta h_{EO}$ = ±10 %, but this is far greater than the actual uncertainty in thickness across the dimensions of a small chip like the ones used in this work. As an example, the BTO used in this work had a thickness of 485 nm. Since the BTO was grown epitaxially and its unit cell is roughly 0.4 nm, the typical error in thickness is on the order of just a few nanometers, which is less than 1 % of the total film thickness. We therefore conclude that variation of $h_{EO}$ in this type of phase shifter only influences the measured results by roughly 1 % at worst.

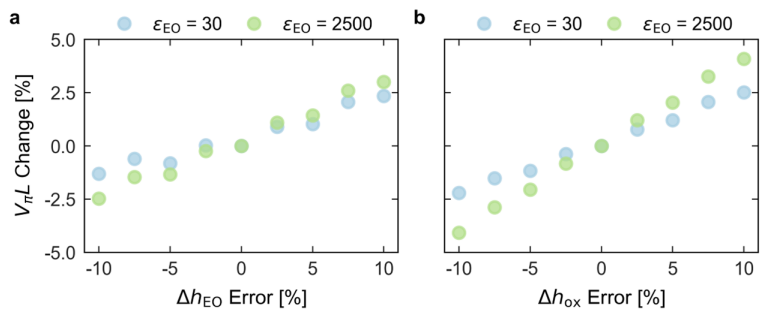

**Fig. S11 | Influence of layer thicknesses on the voltage-length product.** Simulated dependence of the voltage-length length product on the thickness of the **a** EO layer and **b** oxide spacer layer for the phase shifters in this work. The thicknesses are expressed as a percentage error from the designed value and the voltage-length products are expressed as a percentage change from the value at the designed thickness (i.e. $\Delta h$ = 0). The dependence is given for two values of $\varepsilon_{EO}$ roughly corresponding to the values for LN (30, blue) and BTO at low frequencies (2500, green).

The only other layer thickness with any non-negligible influence is that of the oxide layer $h_{ox}$ between the EO layer and the electrodes. This, however, is even more inconsequential than variation in $h_{EO}$ as shown in Fig. S11(b). The effect of $\Delta h_{ox}$ follows the same trend as that of $\Delta h_{EO}$. This is because the RF electric field is typically stronger in the EO layer for smaller $h_{ox}$ which leads to smaller $V_\pi L$. The oxide layers in this work were deposited with PECVD after calibrating the deposition rate such that the error in thickness was less than 5 nm for 100 nm of oxide. We can similarly conclude that variation of $h_{ox}$ has an influence on the measured results of less than a few percent.